\def\e{{\rm e}}
\def\del{\partial}
\def\half{{1\over2}}

\def\abs#1{{\left|{#1}\right|}}
\def\vev#1{\langle #1 \rangle}
\def\e{{\rm e}}
\def\del{\partial}
\def\dslash{\del\kern-0.55em\raise 0.14ex\hbox{/}}

\def\Fig{Fig.~\the\figno\nfig}
\newcommand{\PRD}[3]{Phys. Rev. {\bf D{#1}} (19{#2}) {#3}}
\newcommand{\PRL}[3]{Phys. Rev. Lett. {\bf {#1}} (19{#2}) {#3}}
\newcommand{\NPB}[3]{Nucl. Phys. {\bf B{#1}} (19{#2}) {#3}}
\newcommand{\PLB}[3]{Phys. Lett. {\bf B{#1}} (19{#2}) {#3}}
\newcommand{\PTP}[3]{Prog. Theor. Phys. {\bf {#1}} (19{#2}) {#3}}
\newcommand{\ANN}[3]{Ann. Phys. {\bf {#1}} (19{#2}) {#3}}

\newcommand{\RPP}[3]{Rep. Prog. Phys. {\bf {#1}} (19{#2}) {#3}}
\newcommand{\PRS}[3]{Proc. R. Soc. London. {\bf A{#1}} (19{#2}) {#3}}
\newcommand{\SJN}[3]{Sov. J. Nucl. Phys. {\bf {#1}} (19{#2}) {#3}}

\documentstyle[12pt,epsf]{article}
\begin{document}
\date{} 
\begin{titlepage}
\begin{flushright}
{KOBE-TH-99-07\\ YITP99-56\\ NBI-HE-99-51}
\end{flushright}
\vspace{2cm}
\begin{center}
{\Large A New Mechanism of Spontaneous SUSY Breaking}
\vskip1.0truein
{\large Makoto Sakamoto}$^{(a)}$
\footnote{E-mail:{\tt sakamoto@oct.phys.sci.kobe-u.ac.jp}},
{\large Motoi Tachibana}$^{(b)}$ \
\footnote{E-mail:{\tt motoi@yukawa.kyoto-u.ac.jp}~~JSPS research
fellow}\\
\vspace*{2mm}and \vspace*{2mm}\\
{\large Kazunori Takenaga}$^{(c)}$
\footnote{E-mail:{\tt takenaga@alf.nbi.dk}}
\vskip0.2truein
\centerline{$^{(a)}$ {\it Department of Physics,
Kobe University, Rokkodai, Nada, Kobe 657-8501, Japan }}
\vspace*{2mm}
\centerline{$^{(b)}$ {\it Yukawa Institute for Theoretical
Physics, Kyoto University, Japan}}
\vspace*{2mm}
\centerline{$^{(c)}$ {\it The Niels Bohr Institute, Blegdamsvej 17,
DK-2100,
Copenhagen $\phi$, Denmark}}
\end{center}
\vskip0.5truein
\centerline{\bf Abstract}
\vskip0.3truein
We propose a new mechanism of spontaneous
supersymmetry breaking. The existence
of extra dimensions with nontrivial topology plays an important role.
We investigate new features
resulted from the mechanism in two simple
supersymmetric $Z_2$ and $U(1)$ models.
One of remarkable features is that there exists a phase
in which the translational invariance for the compactified directions
is broken spontaneously, accompanying the breakdown of the
supersymmetry.
The mass spectrum of the models appeared in reduced dimensions is a
full of variety, reflecting the highly
nontrivial vacuum structure of the models. The Nambu-Goldstone
bosons (fermions) associated with breakdown of symmetries are found
in the mass spectrum.
Our mechanism also yields quite different vacuum structures if
models have different global symmetries.
\vskip0.13truein
\baselineskip=0.5 truein plus 2pt minus 1pt
\baselineskip=18pt
\vskip 2cm
\end{titlepage}
\addtolength{\parindent}{2pt}
\newpage
\section{INTRODUCTION}
It has been considered that the (super)string
theories \cite{gsw} (and/or more fundamental theory such as M-theory)
are
plausible candidates
to govern physics at the Planck scale. In general these theories are
defined in higher space-time dimensions than $4$ because of the
consistency of the theories. \par
In the region of enough low energies, however, we have already known
that our space-time is $4$-dimensions, so that extra dimensions must be
compactified by a certain mechanism \cite{kaluza}, and
the supersymmetry (SUSY), which
is usually possessed by those theories, has to be broken because it is
not
observed in the low energy region.
At this stage, mechanisms of compactifications and SUSY
breaking are not fully understood.
\par
It may be interesting to consider quantum field theory in space-time
with one of spaces being multiply connected. This is because
we may shed new light on unanswered questions
in the (supersymmetric) standard model and/or expect
some new dynamics which is useful to seek and understand
new physics beyond it.
Actually, it has been reported that the flavour-blind SUSY breaking
terms can be induced through compactifications by taking account of
possible topological effects of
multiply connected space \cite{ss, fi2, takenaga}.
Therefore, it is important to investigate the physics
which must be possessed by quantum field theory
considered in such the space-time.
\par
We shall consider supersymmetric field theories in
space-time with one of the space coordinates being
compactified. One has to specify
boundary conditions of fields for the compactified directions
when the compactified space is
multiply connected. Contrary to the finite
temperature field theory, we
do not know {\it a priori} what they should be.
We shall relax the conventional periodic boundary condition
to allow nontrivial boundary
conditions on the fields \cite{hosotani,isham}.
If the compactified space is topologically
nontrivial, the idea of the nontrivial boundary conditions
is naturally realized. Due to the topology of the space, the
configuration
space also has nontrivial topology. This causes
an ambiguity in quantization of the theory on the space to
yield an undetermined parameter in the theory \cite{rs}.
The boundary conditions of the fields are twisted with the parameter.
\par
We shall propose a new mechanism of the spontaneous SUSY
breaking through compactification. Extra dimensions
with nontrivial topology
and nontrivial boundary conditions for the extra dimensions
play a central role. Our mechanism is quite
different from the known mechanisms such as
the O'Raifeartaigh \cite{or}
and Fayet-Iliopoulos ones \cite{fi}.
A key point of our mechanism is that
solutions to ${\it F}$-term conditions exist but they
are not realized as vacuum
configurations because of the nontrivial boundary conditions.
\par
As a remarkable consequence of our mechanism,
there appears a nontrivial phase structure with respect to
the size of the compactified space. Namely, the
translational invariance for the compactified directions
is broken spontaneously \cite{djackiw, isham2},
accompanying the
SUSY breaking when the size of the compactified
space exceeds a certain critical value. The curious
vacuum structures resulted from the
mechanism also have an influence on the mass spectrum of the theory.
Depending on the size of the compactified space,
we have a mass spectrum full of variety.
The mass spectrum includes Nambu-Goldstone bosons (fermions)
corresponding to the breakdown of the global symmetries of the theory.
\par
In previous two papers, we presented our
idea \cite{stt1} and applied it to the simple
supersymmetric $Z_2$ model \cite{stt2}.
In those papers, however, the details have been omitted and,
in particular, the mass spectrum of the model has been
discussed very shortly. Since the mass spectrum has a quite
nontrivial dependence on the size of the compactified
space, it is worth investigating it more thoroughly.
In this paper, we shall study the mass spectrum of the $Z_2$ model
in detail, including general discussions of our new mechanism.
We shall also study a new model
called $U(1)$ model in order to see how the
mechanism works and to study new features different from
those of the $Z_2$ model.
\par
In the section $2$ we will discuss a key idea of
our mechanism and study a general structure of
superpotentials realizing our new mechanism.
In the sections $3$ and $4$ we will study
two minimal models called $Z_2$ and $U(1)$ models in detail.
The vacuum structures are determined
and the mass spectrum of the models
appeared in reduced dimensions are analyzed.
In particular, we pay attention to the
Nambu-Goldstone bosons and fermions
associated with the breakdown of global symmetries, such as the
translational invariance for the compactified direction and the
supersymmetry. The section $5$ is devoted to conclusions and
discussions.
We summarize mathematical tools in Appendices
${\rm A}$ and ${\rm B}$ which are
used in the studies of the sections $3$ and $4$.
\section{GENERAL DISCUSSION OF OUR MECHANISM}
In this section we shall discuss
a basic idea of our SUSY breaking mechanism, and
clarify how to construct supersymmetric models in which
our mechanism works. Two minimal models, called $Z_2$ and
$U(1)$ models, will be given explicitly.
\par
Let $W(\Phi)$ be a superpotential consisting of the
chiral superfields $\Phi_j$. The scalar potential is then given by
\begin{equation}
V(A)=\sum_j\abs{F_j}^2=\sum_j\abs{\frac{\del W(A)}{\del A_j}}^2,
\label{supot}
\end{equation}
where $A_j(F_j)$ denotes the lowest (highest) component of $\Phi_j$.
The supersymmetry would be unbroken if there are
solutions to the ${\it F}$-term conditions
\begin{equation}
-F_j^*=\frac{\del W(A)}{\del A_j}\bigg|_{A_k = \bar{A}_k} = 0 \qquad
{\rm for~~all} \quad j,
\label{fterm1}
\end{equation}
since the solutions lead to $V(\bar{A})=0$. Our idea of the
supersymmetry breaking is simple: We impose nontrivial boundary
conditions on the superfields for compactified directions.
They must be consistent with the single-valuedness of the
Lagrangian but inconsistent with the ${\it F}$-term conditions
(\ref{fterm1}). Then, any solution to the ${\it F}$-term conditions
will not be realized as a vacuum configuration of the model.
Thus, we expect that $V(\langle A \rangle) > 0$ because
$\langle A_j \rangle \neq \bar{A_j}$ and that the supersymmetry
is broken spontaneously. We should notice that our mechanism
is obviously different from the O'Raifeartaigh one \cite{or}
since solutions to the ${\it F}$-term conditions are allowed
to exist in our mechanism.
\par
In order to realize such the mechanism, let us
consider the theory in space-time with
one of the space coordinates, say, $y$ being compactified on a circle
$S^1$ whose radius is $R$. Since $S^1$ is multiply connected, we have to
specify boundary conditions for the $S^1$ direction. Let us
impose nontrivial boundary conditions on superfields defined by
\begin{equation}
\Phi_j(x^{\mu}, y+2\pi R) = e^{2\pi i \alpha_j}\Phi_j(x^{\mu}, y).
\label{bccond}
\end{equation}
The phase $\alpha_j$ should be the one such that the Lagrangian density
is
single-valued, {\it i.e.}
\begin{equation}
{\cal L}(x^{\mu}, y+2\pi R) = {\cal L}(x^{\mu}, y).
\label{lagra}
\end{equation}
In other words, the phase has to be
one of the degrees of freedom of
symmetries of the theory. Suppose that $\bar{A_j}$, which is a
solution to the ${\it F}$-term conditions, is a nonzero
constant for some $j$. It is easy to see that if
\begin{equation}
e^{2\pi i \alpha_j} \neq 1, \qquad {\rm for~~some} \quad j,
\label{nontbc}
\end{equation}
then, the vacuum expectation value $\vev{A_j}$ is strictly forbidden
to take the nonzero constant $\bar{A_j}$ because it is
inconsistent with the boundary condition (\ref{bccond}).
In this way, our idea is realized by the mechanism that
the nontrivial boundary conditions imposed
on the fields play a role to force
the vacuum expectation values of the fields not to
take the solutions to the ${\it F}$-term conditions.
The mechanism clearly does not depend
on the space-time dimensions,
so that it would work in any dimensions.
\par
Note that the above result does not always lead us to a
conclusion that $\vev{A_j}= 0$, which is always consistent
with (\ref{bccond}).
Certainly, if the translational invariance for the $S^1$ direction is
not broken, $\vev{A_j}$ has to vanish because
of (\ref{bccond}) with (\ref{nontbc}).
If the translational invariance for the $S^1$ direction
is broken, however, vacuum expectation values will
no longer be constants and some of $\vev{A_j}$'s can depend on
the coordinate of the compactified space as an energetically
favourable configuration. One should include
the contributions from the kinetic terms of the scalar fields
to the scalar potential in order to find the true vacuum configuration.
\par
Here, it is worth making comments on
the nontrivial boundary conditions (\ref{bccond}) imposed
on the fields $\Phi_j$. The boundary conditions have to be one of the
symmetry degrees of freedom of the theory in order to maintain the
single-valuedness of the Lagrangian. One might think that
the $U(1)_R$ symmetry, which is specific to
the supersymmetric theory, is
available. It is well-known that
the boundary condition associated with
the $U(1)_R$ symmetry ${\it explicitly}$ breaks the
supersymmetry \cite{ss, fi2}. In this paper
we do not consider this case, though it is known to have attractive
features \cite{takenaga}. The boundary conditions (\ref{bccond})
we consider in this paper are
consistent with the supersymmetry.
When space is multiply connected and hence has nontrivial
topology, the configuration space can also have nontrivial
topology. Then, it turns out that undetermined parameters
like $\alpha_j$'s in (\ref{bccond}) inevitably appear as the
ambiguity in quantizing the theory on such the space and the
boundary conditions are twisted according to the parameters
\cite{rs}.
Therefore, even though we state
to impose the nontrivial boundary conditions on the fields here, such
the
boundary conditions are firm consequences when one studies the
theory on multiply connected space.
\par
Now, let us clarify a general structure of superpotentials
in which our mechanism works. Let us restrict ourself to $4$-dimensions
for the illustration. According to the
previous discussions, we see that the scalar potential has to be
of the usual Higgs-type to exclude the origin of $A_j=0$ as a
supersymmetric vacuum. So, we require the superpotential to satisfy
\begin{equation}
\frac{\del W(A)}{\del A_j}\bigg|_{A_k =0} \neq 0
\qquad {\rm for~~some} \quad j.
\label{equc}
\end{equation}
It immediately follows that $W(A)$ should
contain {\it linear terms} with respect to $A_j$, {\it i.e.}
\begin{equation}
W(A) = \sum_{j}\lambda_j A_j + \cdots,
\label{equd}
\end{equation}
where some of $\lambda_j$ are nonzero constants.
First, let us consider a model consisting of only one chiral
superfield. From (\ref{equd}), the superpotential is of the form
\begin{equation}
W(\Phi) = \lambda \Phi + f(\Phi),
\label{eque}
\end{equation}
where $f(\Phi)$ stands for higher order terms.
We can immediately conclude that our mechanism does not work for the
superpotential (\ref{eque}) due to the linear
term $\lambda \Phi$. This is because the
superpotential cannot possess any
symmetries which are consistent with the nontrivial boundary
condition of the field.
The phase has to be trivial $e^{2\pi i\alpha}=1$ in this case
in order to keep the single-valuedness of the Lagrangian.
\par
Let us next consider a model with two chiral superfields
$\Phi_0$ and $\Phi_1$. The superpotential is written, taking account of
the requirement (\ref{equc}), as
\begin{equation}
W(\Phi_0,\Phi_1) = \lambda_0 \Phi_0 + \lambda_1 \Phi_1+g(\Phi_0,\Phi_1),
\end{equation}
where $g(\Phi_0,\Phi_1)$ stands for higher order interactions.
If both $\lambda_0$ and $\lambda_1$ are nonzero, the superpotential
does not again possess any symmetries, so that we cannot impose
nontrivial boundary conditions consistent with the single-valuedness
of the Lagrangian. Therefore, one
of them, say $\lambda_1$, has to be zero, {\it i.e.}
\begin{equation}
W(\Phi_0,\Phi_1) = \lambda_0 \Phi_0 +g(\Phi_0,\Phi_1).
\end{equation}
In order to maintain the single-valuedness
of the Lagrangian, $\Phi_0$ has to obey a periodic boundary condition
$\Phi_0(x^{\mu},y+2\pi R) = \Phi_0(x^{\mu},y)$. Thanks to the two
chiral superfields, $\Phi_1$ can have a nontrivial boundary condition.
We have arrived at a possible form of
the superpotential, written up to cubic interactions, as
\begin{eqnarray}
W(\Phi_0,\Phi_1)&=& \lambda_0 \Phi_0 + m_0(\Phi_0)^2+ m_1(\Phi_1)^2
+ m_{01}\Phi_0 \Phi_1 \nonumber\\
&+& g_1(\Phi_0)^3 + g_2(\Phi_0)^2 \Phi_1 + g_3\Phi_0(\Phi_1)^2
+ g_4(\Phi_1)^3.
\end{eqnarray}
This superpotential can have a $Z_2$ or $Z_3$ symmetry, depending on
how we choose the parameters ($\lambda_0$, $m_0$, $\cdots$ , $g_4$).
Boundary conditions of the fields have to be consistent with
the symmetry. The superpotential with the $Z_2$ symmetry is written as
\begin{equation}
W(\Phi_0,\Phi_1)_{Z_2} = \lambda_0 \Phi_0 + m_0(\Phi_0)^2 +m_1(\Phi_1)^2
+ g_1(\Phi_0)^3 + g_3\Phi_0(\Phi_1)^2,
\label{equf}
\end{equation}
which is invariant under
$\Phi_0 \rightarrow \Phi_0$ and $\Phi_1 \rightarrow -\Phi_1$.
On the other hand, the superpotential
\begin{equation}
W(\Phi_0,\Phi_1)_{Z_3} = \lambda_0 \Phi_0 + m_0(\Phi_0)^2
+ g_1(\Phi_0)^3 + g_4(\Phi_1)^3.
\label{equg}
\end{equation}
has the $Z_3$ symmetry whose transformation is given by
$\Phi_0\rightarrow \Phi_0$ and $\Phi_1\rightarrow e^{2i\pi/3}\Phi_1$.
\par
We shall first concentrate on the $Z_2$ symmetric case and
restrict further the form of the
superpotential (\ref{equf}) in order for our
mechanism to work.
Since $W(\Phi_0,\Phi_1)_{Z_2}$ has the $Z_2$ symmetry
we can impose the  nontrivial boundary conditions
on $\Phi_1$ defined by
\begin{equation}
\Phi_0(x^{\mu}, y+2\pi R) = +\Phi_0(x^{\mu}, y), \quad
\Phi_1(x^{\mu}, y+2\pi R) = -\Phi_1(x^{\mu}, y).
\label{equh}
\end{equation}
The next task we have to do is to restrict
the form of the superpotential (\ref{equf})
in such a way that there exists a nontrivial solution to
the ${\it F}$-term conditions
\begin{equation}
\frac{\del W(A_0,A_1)_{Z_2}}{\del A_j}
\bigg|_{A_0=\bar{A}_0,A_1=\bar{A}_1} = 0, \qquad {\rm with}\qquad
\bar{A_1} \neq 0.
\label{equi}
\end{equation}
The $F$-term conditions for the case are
\begin{eqnarray}
0=\frac{\del W_{Z_2}}{\del A_0}\bigg|_{A_0=\bar{A}_0,A_1=\bar{A}_1}
&=& \lambda_0+2m_0\bar{A_0}
+ 3g_1\bar{A_0}^2 + g_3\bar{A_1}^2, \nonumber\\
0=\frac{\del W_{Z_2}}{\del A_1}\bigg|_{A_0=\bar{A}_0,A_1=\bar{A}_1}
&=& 2\bar{A_1}(m_1 + g_3 \bar{A_0}).
\label{equjcond}
\end{eqnarray}
It is easy to see that the condition $\bar{A_1} \neq 0$ requires at
least
$g_3 \neq 0$. Then, the second condition in (\ref{equjcond}) leads to
\begin{equation}
\bar{A_1} =  0 \quad {\rm or} \quad m_1 + g_3\bar{A_0} = 0.
\label{equkk}
\end{equation}
The first solution $\bar{A_1} = 0$, which is actually unwanted, can
be excluded by requiring
$m_0 = g_1 = 0$ and $\lambda_0\neq 0$.
With these choices of the parameters, $\bar{A_1} = 0$ is no longer
a solution of (\ref{equjcond})
and the second solution in (\ref{equkk}) is selected.
As the result, the solutions of
the ${\it F}$-term conditions are given by
$(\bar{A_0}, \bar{A_1})=(-\frac{m_1}{g_3},
\pm \sqrt{-\frac{\lambda_0}{g_3}})$.
The solutions, however, cannot be realized as vacuum
configurations because they are inconsistent with the boundary
conditions (\ref{equh}). Thus, we have found the superpotential
which satisfies our criteria
(\ref{equc}) and (\ref{equi}) to realize our mechanism to be
of the form
\begin{equation}
W(\Phi_0,\Phi_1)_{Z_2} = \lambda_0\Phi_0 + m_1(\Phi_1)^2
+ g_3\Phi_0(\Phi_1)^2.
\label{equl}
\end{equation}
Let us note that by shifting $\Phi_0$ by $-m_1/g_3$, we can
also rewrite (\ref{equl}) as
\begin{equation}
W(\Phi_0,\Phi_1)_{Z_2} = \lambda_0\Phi_0+ g_3\Phi_0(\Phi_1)^2
\label{equm}
\end{equation}
up to constant.
This is a minimal model to realize our mechanism.
Let us note that the superpotential ({\ref{equm}) has another global
symmetry $U(1)_R$ in addition to the $Z_2$ symmetry.
As we stated earlier, we
do not consider any boundary conditions
associated with the $U(1)_R$ symmetry degrees of freedom
in this paper, concerning about the boundary conditions.
\par
It is easy to see that the scalar potential followed from the
superpotential (\ref{equl}) or (\ref{equm}) is the usual
Higgs-type potential. It is essential for our mechanism that
the scalar potential is the Higgs-type to have the minimum
of the potential away from the origin and that nonvanishing
$\bar{A_1}$, which is any solution to the ${\it F}$-term conditions,
is inconsistent with the boundary condition of $A_1$.
\par
Let us come back to the superpotential (\ref{equg})
 with the $Z_3$ symmetry.
In this case, $\Phi_0$ decouples with $\Phi_1$ completely.
It is easy to see that the ${\it F}$-term condition
for $A_1$ inevitably gives us $\bar{A_1}=0$ for {\it any}
choice of the parameters in the
superpotential, so that our mechanism does not work in this case.
\par
We can also construct a supersymmetric model
with a ${\it continuous}$ symmetry.
In this case the minimal model is given by introducing at
least ${\it three}$ chiral superfields.
The superpotential with a global $U(1)$ symmetry is given by
\begin{equation}
W(\Phi_0,\Phi_1)_{U(1)} = \lambda_0\Phi_0 + g\Phi_0 \Phi_+ \Phi_-
+m\Phi_+\Phi_-
\label{equn}
\end{equation}
with the boundary conditions
\begin{equation}
\Phi_0(x^{\mu}, y+2\pi R) = \Phi_0(x^{\mu}, y), \quad
\Phi_{\pm}(x^{\mu}, y+2\pi R) = e^{\pm 2\pi i\alpha}\Phi_{\pm}(x^{\mu},
y).
\label{equo}
\end{equation}
The $U(1)$ charges of $\Phi_0$ and $\Phi_{\pm}$ are
defined by 0 and $\pm 1$, respectively.
By shifting $\Phi_0$ by $-m/g$ in (\ref{equn}), it is also written as
\begin{equation}
W(\Phi_0,\Phi_1)_{U(1)} = \lambda_0\Phi_0 + g\Phi_0 \Phi_+ \Phi_-
\label{equp}
\end{equation}
up to constant.
Here, this superpotential has the $U(1)_R$ symmetry, in addition to the
global $U(1)$ symmetry.
\par
Through the discussions we have made, we finally find
the conditions on the superpotential to be satisfied
in order for our mechanism to work:
\begin{itemize}
\item The origin of $A_j=0$ is not
the supersymmetric vacuum of the potential $V(A)$.
In other words, $A_j=0$ for all $j$ is not a solution
to the ${\it F}$-term conditions.
\item Let $\bar{A_j}$ be a configuration which
satisfies the ${\it F}$-term conditions.
Some of $A_j$ with $\bar{A_j}\neq 0$ have to be non-singlets for some
global symmetries of the theory.
\end{itemize}
A key point of our mechanism is that the non-singlet
fields $A_j$ with $\bar{A_j}\neq 0$ are
required to obey nontrivial boundary
conditions consistent with the single-valuedness of the Lagrangian.
It should be emphasized that
solutions of the ${\it F}$-term conditions
always exist in our case but are not realized as vacuum configurations
due to the nontrivial boundary conditions.
Thus, our mechanism is different from other ones
of the spontaneously SUSY breaking, {\it e.g.}
O'Raifeartaigh models, in which
there are no consistent solutions of the $F$-term conditions.
In the following two sections, we shall study the $Z_2$
and $U(1)$ models whose superpotentials are given by
(\ref{equl}) ( or (\ref{equm}))
and (\ref{equn}) ( or (\ref{equp})), respectively, in detail.
\section{THE ${\bf Z_2}$ MODEL}
In this section we study the $Z_2$ model obtained
in the previous section. In particular, we study the vacuum structure
and the mass spectrum for bosons and fermions of the model.
\subsection{Vacuum configuration}
The superpotential of the $Z_2$ model is given by
\begin{equation}
W(\Phi_0,\Phi_1) = g\Phi_0 \left[ \frac{\Lambda^2}{g^2}-
\frac{1}{2}(\Phi_1)^2\right] +\frac{\mu}{2}(\Phi_1)^2.
\label{z2w}
\end{equation}
We have slightly changed notations of the
superpotential (\ref{equl}) in the previous section.
The parameters $g$ and $\Lambda$ are
complex in general, but their phases can be absorbed into the
redefinitions of the fields $\Phi_0$ and $\Phi_1$. Thus, we
take $g$ and $\Lambda$ to be real without loss of generality.
The model at hand is invariant under a discrete $Z_2$ transformation
$\Phi_0\rightarrow \Phi_0, \Phi_1\rightarrow -\Phi_1$.
As discussed in the previous section,
the existence of the global symmetry is crucial
for our mechanism because otherwise nontrivial boundary
conditions cannot be imposed on the fields.
The scalar potential is given by
\begin{equation}
V(A_0,A_1) \equiv |F_0|^2 + |F_1|^2
= \bigg|\frac{\Lambda^2}{g} -
\frac{g}{2}(A_1)^2\bigg|^2 + |gA_0 - \mu|^2 |A_1|^2.
\label{z2v}
\end{equation}
It is easy to solve the
${\it F}$-term conditions, and the solutions are given by
\begin{equation}
\bar{A_0}=\mu/g, \qquad \bar{A_1}=\pm{{\sqrt{2}\Lambda}/g},
\label{fsol}
\end{equation}
at which $V(\bar{A_0}, \bar{A_1})$ vanishes. Therefore, one might think
that the supersymmetry is unbroken while the $Z_2$ symmetry is broken
spontaneously. This is, however, a hasty conclusion,
as we will see below.
\par
Let us consider the model on $M^3\otimes S^1$. We shall denote the
coordinates of $M^3$ and $S^1$
 by $x^{\mu}(\mu=0,1,2)$ and $y$, respectively.
Since $S^1$ is multiply connected,
we can impose the nontrivial boundary
conditions on the superfields
\begin{equation}
\Phi_0(x^{\mu},y+2\pi R) = +\Phi_0(x^{\mu},y), \quad
\Phi_1(x^{\mu},y+2\pi R) = -\Phi_1(x^{\mu},y).
\label{zbc}
\end{equation}
It follows that the vacuum expectation value of $A_1$
(and also the auxiliary field $F_1$) is forced not to
take a nonzero constant, {\it i.e.} $\langle A_1(x^{\mu},y) \rangle
\neq$ nonzero constant. Therefore, the solutions (\ref{fsol}) to the
${\it F}$-term conditions are not consistent with the boundary
conditions (\ref{zbc}) and hence are not realized as vacuum
configurations. Since any supersymmetric vacuum has to satisfy the
${\it F}$-term conditions, if any, and since all the solution to the
${\it F}$-term conditions are excluded from vacuum configurations
by the boundary conditions, the supersymmetry is force to be
broken spontaneously, so that our mechanism does work in this model.
\par
A configuration consistent with the boundary conditions (\ref{zbc}) is
$\vev{A_0}=\mu/g, \vev{A_1}=0$. One may wonder whether
this is a vacuum configuration or not. The configuration
seems unstable because the scalar potential, which is
now the Higgs-type, has a negative curvature along the
$A_1$ direction at $\vev{A_0}=\mu/g$.
When we search vacuum configurations, we usually minimize the scalar
potential. This will not, however, lead to the true vacuum
configuration in the model we are considering. Since one of the space
coordinates is compactified on $S^1$, the original fields may be
decomposed into Kaluza-Klein modes for the $S^1$ direction. The
kinetic term for the $S^1$ direction can be regarded as mass terms
for the Kaluza-Klein modes from the 3-dimensional point of view.
An important observation is that all the Kaluza-Klein modes for the
field $A_1$ acquire nonvanishing masses since $A_1(y)$ obeys the
antiperiodic boundary condition and has no zero mode. This suggests
that the kinetic term of $A_1$ for the $S^1$ direction could
drastically change the shape of the potential (in the 3-dimensional
point of view) near the origin $A_1=0$ (recall that $V(A_0,A_1)$ is
the Higgs-type). Therefore, we should study the scalar potential
in which the contribution from the kinetic terms for the $S^1$
direction is taken into account in order to
find the true vacuum configuration.
\par
According to the above discussions, the vacuum configuration will be
obtained by solving a minimization
problem of the ``energy'' functional
\begin{equation}
{\cal E}[A_0, A_1; R] \equiv \int_{0}^{2\pi R} dy\left[
\bigg|\frac{dA_0}{dy}\bigg|^2 +
\bigg|\frac{dA_1}{dy}\bigg|^2 + V(A_0, A_1) \right ]
\label{efunc}
\end{equation}
with the boundary conditions
\begin{equation}
A_0(y+2\pi R) = +A_0(y), \quad
A_1(y+2\pi R) = -A_1(y).
\label{zsbc}
\end{equation}
We assume that the translational invariance for the
$3$-dimensional Minkowski space-time is not broken,
so that we ignore the $x^{\mu}$-dependence of $A_0$ and $A_1$.
\par
The detailed discussions on the minimization of the
``energy'' functional (\ref{efunc}) are made in the Appendix A. Let us
present the results here. We obtain the vacuum configuration as
\begin{eqnarray}
&{\rm for}&R \leq R^*\equiv
\frac{1}{2\Lambda}\quad \left\{\begin{array}{ll}
\vev{A_0(x^{\mu},y)} =& {\rm arbitrary~complex~constant},\\
\vev{A_1(x^{\mu},y)} =& 0,  \\
\end{array}\right. \label{zvac1}\\
&{\rm for}&R>R^*\equiv \frac{1}{2\Lambda}\quad \left\{\begin{array}{ll}
\vev{A_0(x^{\mu},y)} =& \mu/g, \\
\vev{A_1(x^{\mu},y)} =& \frac{2k\omega}{g}{\rm sn}(\omega(y-y_0), k).
\end{array}\right.
\label{zvac2}
\end{eqnarray}
Here, the parameter $k~(0\le k <1)$ and $y_0$
are the integration constants
and $\omega \equiv \frac{\Lambda}{\sqrt{1+k^2}}$.
The ${\rm sn}(u,k)$ is the Jacobi elliptic function whose period is
$4K(k)$,
where $K(k)$ denotes the complete elliptic function of the first kind.
Since the integration constant $y_0$, which reflects the translational
invariance of the equation of motion, is irrelevant, so that
hereafter we set $y_0=0$.
\par
We observe a remarkable feature
that in our model there are two phases, depending on
the magnitude of the radius of $S^1$.
We see that the vacuum structure of the
model drastically changes at the critical radius $R^*$.
For $R \leq R^*$,
the translational invariance for the $S^1$ direction
and the $Z_2$ symmetry are unbroken,
while for $R > R^*$, since $\vev{A_{1}(x^{\mu},y)}$ depends on the
coordinate $y$ of the extra dimension, the
translational invariance for the $S^1$ direction is broken
spontaneously with the breakdown of the $Z_2$ symmetry.
The vacuum energy is nonzero for both
$R \leq R^*$ and $R > R^*$, so that
the SUSY is broken spontaneously for both regions.
\par
As discussed in the Appendix A, the parameter $k~(0\le k <1)$
is determined through the condition
\begin{equation}
2\pi R=\frac{2}{\Lambda}\sqrt{1+k^2}K(k)
\label{zconst}
\end{equation}
in order for (\ref{zvac2}) to satisfy the boundary
conditions (\ref{zsbc}). Knowing that $K(k)$
is a monotonically increasing function of $k$, we immediately obtain
the critical radius $R^*\equiv 1/{2\Lambda}$, which
corresponds to $k=0$.
Though we present the detailed discussion on the critical radius
in the Appendix A, it may be instructive to show how the critical
radius $R^*$ appears in the 3-dimensional point of view. To this end,
we may expand $A_0(x^{\mu}, y)$ and $A_1(x^{\mu}, y)$ in the Fourier
modes for the $y$ direction, according to the boundary conditions
(\ref{zsbc}). As found in the next subsection (see also the Table
III-1),
the squared mass eigenvalues for the Fourier modes of $A_0$ are
given by $(\frac{n}{R})^2$ and those for $A_1$ by
$|\mu - g\langle A_0  \rangle|^2 \pm \Lambda^2 +
(\frac{n+\frac{1}{2}}{R})^2$ with $n$ being integers.
It follows that that for $R \leq \frac{1}{2\Lambda}$ all the squared
masses are positive semi-definite and hence the vacuum configuration
(\ref{zvac1}) is stable at least locally, while for
$R > \frac{1}{2\Lambda}$ negative squared masses would appear with
$\langle A_0  \rangle = \frac{\mu}{g}$. This implies that the
configuration $\langle A_1  \rangle = 0$ is unstable and can
no longer be a vacuum for $R > \frac{1}{2\Lambda}$. The phase
transition should occur at $R = R^* = \frac{1}{2\Lambda}$.
As proved in the Appendix A, the vacuum configuration becomes
stable by having the coordinate dependence such as (\ref{zvac2}).

\subsection{Mass spectrum for $R \leq R^*$}
Once the vacuum configurations are determined,
we shall next analyze
the mass spectrum of the model. We shall employ the standard
prescription
based on perturbation theory to obtain the mass spectrum.
We expand fields around
the vacuum configuration and take the quadratic terms
with respect to the fluctuations. The fluctuating fields have to
be expanded appropriately in modes of
oscillations about the vacuum configuration.
Since we are interested in the mass spectrum
appeared in the $3$-dimensional
Minkowski space-time, we integrate over the coordinate $y$ of
the extra dimension. In this way we can obtain the mass spectrum
for the bosons and fermions in the model.
\par
Let us first compute the bosonic mass spectrum.
The vacuum configuration is given by (\ref{zvac1}) for $R \leq R^*$.
The bosonic mass spectrum can be read from the quadratic terms
of the scalar potential, including the kinetic terms of the scalar
fields for the $S^1$ direction. It is given by
\begin{equation}
{\cal L}_{B(R \le R^*)}^{(2)}=
\int_0^{2\pi R}dy\Bigl[
-\sum_{i=0,1}\del_M A_i^*\del^M A_i +
\frac{\Lambda^2}{2}\bigg( (A_1)^2+(A_1^*)^2 \bigg)
-|M|^2|A_1|^2
\Bigr],
\label{z2scalar}
\end{equation}
where $M \equiv \mu -g\langle A_0 \rangle$.

We find that it
is convenient to introduce $4$ real scalar fields defined by
\begin{equation}
A_{i}(x^{\mu},y) \equiv
\frac{1}{\sqrt{2}}(a_{i}(x^{\mu},y) + i b_{i}(x^{\mu},y)),
\qquad (i=0,1).
\label{4reals}
\end{equation}
Moreover, in order to obtain the mass spectrum appeared in
$3$-dimensions,
we expand $a_0$, $b_0$, $a_1$ and $b_1$ in Fourier
series for the $S^1$ direction, according to
the boundary conditions (\ref{zsbc}), as
\begin{eqnarray}
a_0(x, y) &=& \frac{1}{\sqrt{2\pi R}}a_0^{(0)}(x)
+ \frac{1}{\sqrt{\pi R}}\sum_{n \in{\bf Z}>0}\bigg[a_0^{c(n)}(x)
\cos(\frac{n y}{R}) + a_0^{s(n)}(x)\sin(\frac{n y}{R})\bigg],
\\
b_0(x, y) &=& \frac{1}{\sqrt{2\pi R}}b_0^{(0)}(x)
+ \frac{1}{\sqrt{\pi R}}\sum_{n \in{\bf Z}>0}\bigg[b_0^{c(n)}(x)
\cos(\frac{n y}{R}) + b_0^{s(n)}(x)\sin(\frac{n y}{R})\bigg],
\\
a_1(x, y) &=&
\frac{1}{\sqrt{\pi R}}\sum_{l\in {\bf Z}+\half>0}\bigg[a_1^{c(l)}(x)
\cos(\frac{l y}{R}) + a_1^{s(l)}(x)\sin(\frac{l y}{R})\bigg],
\\
b_1(x, y) &=&
\frac{1}{\sqrt{\pi R}}\sum_{l\in{\bf Z}+\half>0}\bigg[b_1^{c(l)}(x)
\cos(\frac{l y}{R}) + b_1^{s(l)}(x)\sin(\frac{l y}{R})\bigg].
\end{eqnarray}
Note that the Fourier modes of the fields $a_1(x)$ and $b_1(x)$
cannot have zero modes due to the boundary condition (\ref{zsbc}).
Inserting these expressions into (\ref{z2scalar}) and integrating over
the
coordinate $y$, we obtain the mass spectrum of the bosons.
The result is summarized in the Table III-1.
\par
The computations of the fermion mass spectrum can be
done just as in the bosonic case.
The mass spectrum for the fermions can be read from
the quadratic terms given by
\begin{equation}
{\cal L}_{F(R \leq R^*)}^{(2)}=
\int_{0}^{2\pi R}dy [ -i\sum_{i=0,1}\bar{\psi_i}\bar{\sigma}^{M}
\del_{M}\psi_i
-\frac{1}{2}M\psi_1\psi_1
-\frac{1}{2}M^*\bar{\psi_1}\bar{\psi_1}].
\end{equation}
We are interested in the spectrum appeared in $3$-dimensions.
We find that it is convenient to introduce the Dirac spinors in
$3$-dimensions according to our prescription (\ref{diracs}) in
the Appendix B and
to decompose further the Dirac spinors into the Majorana
spinors (\ref{majo}) in $3$-dimensions.
In the Appendix B, we present details of
the decomposition of the 4-dimensional
spinors and gamma matrices into the 3-dimensional ones.
According to these, we obtain
\begin{eqnarray}
{\cal L}_{F(R\leq R^*)}^{(2)}&=&
\int_{0}^{2\pi R}dy \bigg[
-\frac{i}{2}\sum_{i=0,1}\bar{\chi_i}\gamma^{\mu}
\del_{\mu}\chi_i  -\frac{i}{2}\sum_{i=0,1}\bar{\rho_i}\gamma^{\mu}
\del_{\mu}\rho_i  \nonumber \\
& &-\frac{1}{2}\sum_{i=0,1}\bar{\chi_i}\del_y \rho_i
+\frac{1}{2}\sum_{i=0,1}\bar{\rho_i}\del_y \chi_i \nonumber\\
& &+\frac{i}{4}(M-M^*)(\bar{\chi_1}\chi_1-\bar{\rho_1}\rho_1)
-\frac{1}{4}(M+M^*)(\bar{\chi_1}\rho_1+\bar{\rho_1}\chi_1)\bigg],
\label{zfermi}
\end{eqnarray}
where $\chi_0$, $\rho_0$, $\chi_1$ and $\rho_1$ are the
Majorana spinors in $3$-dimensions.
We expand $\chi_0$, $\rho_0$, $\chi_1$ and $\rho_1$ in
Fourier series for the $S^1$ direction as
\begin{eqnarray}
\chi_0(x, y) &=& \frac{1}{\sqrt{2\pi R}}\chi_0^{(0)}(x)
+ \frac{1}{\sqrt{\pi R}}\sum_{n \in{\bf Z}>0}\bigg[\chi_0^{c(n)}(x)
\cos(\frac{n y}{R}) + \chi_0^{s(n)}(x)\sin(\frac{n y}{R})\bigg],
\\
\rho_0(x, y) &=& \frac{1}{\sqrt{2\pi R}}\rho_0^{(0)}(x)
+ \frac{1}{\sqrt{\pi R}}\sum_{n \in{\bf Z}>0}\bigg[\rho_0^{c(n)}(x)
\cos(\frac{n y}{R}) + \rho_0^{s(n)}(x)\sin(\frac{n y}{R})\bigg],
\\
\chi_1(x, y) &=&
\frac{1}{\sqrt{\pi R}}\sum_{l\in{\bf Z}+\half >0}\bigg[\chi_1^{c(l)}(x)
\cos(\frac{l y}{R}) + \chi_1^{s(l)}(x)\sin(\frac{l y}{R})\bigg],
\\
\rho_1(x, y) &=&
\frac{1}{\sqrt{\pi R}}\sum_{l\in{\bf Z}+\half >0}\bigg[\rho_1^{c(l)}(x)
\cos(\frac{l y}{R}) + \rho_1^{s(l)}(x)\sin(\frac{l y}{R})\bigg].
\end{eqnarray}
Let us note again that because of the boundary
conditions (\ref{zbc}) there are no zero
modes in $\chi_1$ and $\rho_1$.
Inserting these expressions into (\ref{zfermi}) and performing the
$y$ integration, we obtain
\begin{eqnarray}
{\cal L}_{F(R \leq R^*)}^{(2)}&=&-\frac{i}{2}{\bar\chi}_0^{(0)}
\gamma^{\mu}\del_{\mu}\chi_0^{(0)}
-\frac{i}{2}{\bar\rho}_0^{(0)}
\gamma^{\mu}\del_{\mu}\rho_0^{(0)}+
\frac{1}{2}\sum_{n\in{\bf Z} >0}
{\bar\Psi}_0^{(n)}
(-i\gamma^{\mu}\del_{\mu}{\bf 1}+{\cal M}_0^{(n)})
\Psi_0^{(n)} \nonumber \\
& & +\frac{1}{2}\sum_{l\in{\bf Z}+\half >0}
{\bar\Psi}_1^{(l)}
(-i\gamma^{\mu}\del_{\mu}{\bf 1}+{\cal M}_1^{(l)})
\Psi_1^{(l)},
\label{eflag}
\end{eqnarray}
where
\begin{eqnarray}
{\cal M}_0^{(n)}&=&
\left(\begin{array}{cccc}
0&0&0&-n/R\\
0&0&n/R&0\\
0&n/R&0&0\\
-n/R&0&0&0
\end{array}\right), \\
{\cal M}_1^{(l)}&=&\left(\begin{array}{cccc}
\frac{i}{2}(M-M^*)&0&-\frac{1}{2}(M+M^*)&-l/R\\
0&\frac{i}{2}(M-M^*)&l/R&-\frac{1}{2}(M+M^*)\\
-\frac{1}{2}(M+M^*)&l/R&-\frac{i}{2}(M-M^*)&0\\
-l/R&-\frac{1}{2}(M+M^*)&0&-\frac{i}{2}(M-M^*)
\end{array}\right), \\ \nonumber
\end{eqnarray}
and $\Psi_0^{(n)}\equiv (\chi_0^{c(n)}, \chi_0^{s(n)}, \rho_0^{c(n)},
\rho_0^{s(n)})^T,
\Psi_1^{(l)}\equiv (\chi_1^{c(l)}, \chi_1^{s(l)}, \rho_1^{c(l)},
\rho_1^{s(l)})^T$.
It may be sufficient to diagonalize, instead of
${\cal M}_{0}^{(n)}$ and ${\cal M}_{1}^{(l)}$,
the square of the each mass matrix
\begin{equation}
({\cal M}_0^{(n)})^2=\bigg(\frac{n}{R}\bigg)^2{\bf 1}_{4\times 4},
\quad
({\cal M}_1^{(l)})^2=\bigg(\abs{M}^2+
\bigg(\frac{l}{R}\bigg)^2\bigg){\bf 1}_{4\times 4}.
\end{equation}
We summarize the fermion mass spectrum in the Table III-2.
\par
\subsection{Analysis of the mass spectrum for $R \leq R^*$}
Let us make several comments on the boson and fermion
mass spectra obtained in the previous subsection.
It is easy to see from the Table III-1 that the boson mass spectrum
is positive semi-definite for $R \leq R^*$.
The vacuum configuration we find is stable for $R \leq R^*$.
The second comment is that the SUSY breaking scale is found,
from the mass splitting, to be of order $\Lambda$. This is also
understood from the fact that the
vacuum energy density for
the vacuum configuration (\ref{zvac1}) is of order $\Lambda$.
The next comment is that we observe that
the modes $A_0^{(0)}\sim a_0^{(0)}+ib_0^{(0)}$ are  massless.
Its physical interpretation is that the massless modes
correspond to the existence of a flat
direction of the scalar potential along the $A_0$ direction
at the tree level. This interpretation is modified slightly
if we start form the
superpotential (\ref{equm}), instead of (\ref{equl}),
 which has an extra $U(1)_R$ symmetry.
The vacuum configuration (\ref{zvac1}) breaks the $U(1)_R$ symmetry
spontaneously because $A_0$ carries the $U(1)_R$ charge. We expect
a Nambu-Goldstone boson associated with the breakdown of
the $U(1)_R$ symmetry to appear. A part of $A_0^{(0)}$ will correspond
to this massless mode and the other part of $A_0^{(0)}$ is also
guaranteed to be massless
by the flatness of the potential at the tree level.
The last comment is that at $\vev{A_0}=\mu/g$, which
yields $M=0$, we find
that the squared masses of some of the bosonic modes $a_1^{c(l)}$
and $a_1^{s(l)}$ become negative for $R > R^* = \frac{1}{2\Lambda}$.
This suggests that the vacuum configuration
becomes unstable for $R > R^*$ and that
a phase transition occurs at $R = R^*$, as discussed in the
previous subsection.
\par
Let us next make comments on the fermion mass spectrum.
We can interpret
$\psi_0^{(0)}\sim \chi_0^{(0)}+i\rho_0^{(0)}$ as the Nambu-Goldstone
fermions associated with the spontaneous SUSY breaking.
In order to confirm this interpretation, let us consider the
infinitesimal SUSY transformations
\begin{eqnarray}
\delta_S A_0(x,y) = \sqrt{2}\xi \psi_0(x,y),\qquad
\delta_S A_1(x,y) = \sqrt{2}\xi \psi_1(x,y), \\ \nonumber
\delta_S \psi_0(x,y) = i\sqrt{2} (\sigma^M \bar{\xi})\partial_M
A_0(x,y) -\sqrt{2}\xi
\left(\frac{\Lambda^2}{g} - \frac{g}{2}(A_1^*(x,y))^2 \right), \\
\delta_S \psi_1(x,y) = i\sqrt{2} (\sigma^M \bar{\xi})\partial_M
A_1(x,y) -\sqrt{2}\xi
\left(\mu - gA_0^*(x,y)\right)A_1^*(x,y).
\label{infsusy}
\end{eqnarray}
Nonvanishing vacuum expectation values
of $\delta_S \psi_0(x,y)$  and $\delta_S
\psi_1(x,y)$ are a signal of the SUSY breaking. They become
\begin{equation}
\langle \delta_S \psi_0(x,y) \rangle = -\sqrt{2}\frac{\Lambda^2}{g}\xi,
\qquad
\langle \delta_S \psi_1(x,y) \rangle = 0
\label{sutra1}
\end{equation}
for the vacuum configuration (\ref{zvac1}).
Since $\langle \psi_0(x, y) \rangle \neq 0$ for any nonvanishing
$\xi$,
the supersymmetry is completely broken spontaneously.
No supersymmetry is left in $3$-dimensions.
Moreover, it follows
from (\ref{sutra1}) that the Nambu-Goldstone modes
associated with the supersymmetry breaking should be
constant modes of $\psi_0$
in Fourier series for
the $S^1$ direction, {\it i.e.}
they are $\chi_0^{(0)}$ and $\rho_0^{(0)}$.
Hence, we confirm the interpretation.
\par
Let us finally note that the mass spectra
for the bosons and fermions satisfy the relations
\begin{eqnarray}
m_{a_0^{(0)}}^2 + m_{b_0^{(0)}}^2 &=&
m_{{\chi}_0^{(0)}}^2 + m_{{\rho}_0^{(0)}}^2, \nonumber \\
m_{a_0^{c(n)}}^2 + m_{a_0^{s(n)}}^2 + m_{b_0^{c(n)}}^2
+ m_{b_0^{s(n)}}^2 &=& m_{\chi_0^{c(n)}}^2+ m_{\chi_0^{s(n)}}^2
+ m_{\rho_0^{c(n)}}^2 + m_{\rho_0^{s(n)}}^2, \nonumber \\
m_{a_1^{c(l)}}^2 + m_{a_1^{s(l)}}^2 + m_{b_1^{c(l)}}^2
+ m_{b_1^{s(l)}}^2 &=& m_{\chi_1^{c(l)}}^2+ m_{\chi_1^{s(l)}}^2
+ m_{\rho_1^{c(l)}}^2 + m_{\rho_1^{s(l)}}^2,
\label{sutramass}
\end{eqnarray}
where $n \in {\bf Z}>0$ and $l \in {\bf Z}+\frac{1}{2}>0$.
These relations mean that the
sum of the squared masses of the bosons for each mode
are exactly equal to that of the fermions.
This is known as the supertrace formula which is an immediate
consequence of the spontaneous SUSY breaking \cite{FGP},
but the relations (\ref{sutramass}) are a stronger version of it.
The formula given in Ref. \cite{FGP} implies the equivalence
between the sum of the squared masses of {\it all} the bosonic
states and that of {\it all} the fermionic states
(in the 3-dimensional point of view), while the relations
(\ref{sutramass}) hold for each mode of the bosonic and
fermionic states.
Before closing this subsection, we shall summarize the mass
spectra of the bosons and fermions for $R \leq R^*$ in the Figure $1$.
\subsection{Mass spectrum for $R > R^*$}
In this subsection, we shall compute the mass spectrum for $R > R^*$.
As we have already mentioned, the vacuum configuration is given
by (\ref{zvac2}) for $R > R^*$.
The vacuum configuration is highly nontrivial
in the sense that it is given by the Jacobi elliptic function which
depends on the coordinate $y$ of the extra dimension  and
the parameter $k$. It means that the vacuum configuration has the
$R$-dependence through the condition (\ref{zconst}).
This also reflects  the mass spectrum, so that it is
interesting to study the $R$-dependence of
the mass spectrum for $R > R^*$.
\subsubsection{Boson sector}
Let us first analyze the mass spectrum of the bosons.
The quadratic terms are given, in terms of the
four real scalar fields (\ref{4reals}), by
\begin{eqnarray}
{\cal L} _{B(R>R^*)}^{(2)}&=& \int_0^{2\pi R} dy \bigg[
-\frac{1}{2}\del_M a_0\del^M a_0
- \frac{1}{2}|g\langle A_1(y)\rangle|^2(a_0)^2 \nonumber\\
&-&\frac{1}{2}\del_M b_0\del^M b_0
- \frac{1}{2}|g\langle A_1(y)\rangle|^2(b_0)^2 \nonumber\\
&-&\frac{1}{2}\del_M a_1\del^M a_1
- \frac{1}{2}\bigg(\frac{3}{2}
|g\langle A_1(y)\rangle|^2-\Lambda^2\bigg)(a_1)^2 \nonumber\\
&-&\frac{1}{2}\del_M b_1\del^M b_1
- \frac{1}{2}\bigg(\frac{1}{2}
|g\langle A_1(y)\rangle|^2+\Lambda^2\bigg)(b_1)^2\bigg].
\label{zboson}
\end{eqnarray}
In order to see how we can obtain the mass spectrum appeared in
$3$-dimensions, let us take a close look at the field $a_0$.
The quadratic terms of the Lagrangian for $a_0$ are read
from (\ref{zboson}) as
\begin{eqnarray}
{\cal L}_{B(R>R^*)}^{(2)}(a_0)&\equiv&\int_0^{2\pi R}dy~\Bigl[
-\frac{1}{2}\del_\mu a_0\del^\mu a_0-\frac{1}{2}(\del_y a_0)^2
-\frac{1}{2}\abs{\vev{gA_1(y)}}^2(a_0)^2\Bigr] \nonumber\\
&=&\int_0^{2\pi R}dy~\Bigl[
-\frac{1}{2}\del_\mu a_0\del^\mu
a_0-\frac{1}{2}a_0{\cal M}_{a_0}^2(y)a_0
\Bigr],
\label{a0bosonf}
\end{eqnarray}
where
\begin{equation}
{\cal M}_{a_0}^2(y)\equiv -\del_y^2+\abs{\vev{gA_1(y)}}^2.
\label{b1op}
\end{equation}
Here, we have carried out the partial integration.
Let us consider the eigenvalue equation defined by
\begin{equation}
{\cal M}_{a_0}^2(y)\phi_{a_0}^{(i)}(y)\equiv
\Bigl[-\frac{d^2}{dy^2}+\abs{g\vev{A_1(y)}}^2\Bigr]
\phi_{a_0}^{(i)}(y)=(m_{a_0}^{(i)})^2\phi_{a_0}^{(i)}(y)
\label{a0boson}
\end{equation}
with the boundary condition
$\phi_{a_0}^{(i)}(y+2\pi R)=\phi_{a_0}^{(i)}(y)$.
Since the set of $\{\phi_{a_0}^{(i)}\}$ is expected to form a complete
set, we may expand $a_0(x^{\mu},y)$ as
$a_0(x^{\mu}, y)=\sum_i a_0^{(i)}(x^{\mu})\phi_{a_0}^{(i)}(y)$
with
$\int_0^{2\pi R}dy \phi_{a_0}^{(i)}(y)\phi_{a_0}^{(j)}(y)=\delta_{ij}$.
Then, (\ref{a0bosonf}) becomes
\begin{equation}
{\cal L}_{B(R>R^*)}^{(2)}(a_0)=\sum_i
\Bigl[-\frac{1}{2}\del_{\mu}a_0^{(i)}\del^{\mu}a_0^{(i)}-
\frac{1}{2}(m_{a_0}^{(i)})^2(a_0^{(i)})^2\Bigr].
\end{equation}
We see that the eigenvalue of the equation
(\ref{a0boson}) is nothing but the
squared mass for $a_0^{(i)}$ in $3$-dimensions.
Therefore, finding the masses of $a_0^{(i)}$ is equivalent to
solving the eigenvalue equation (\ref{a0boson}).
The field $b_0$ satisfies the same equation with $a_0$.
In the same way, we have the eigenvalue
equations for $a_1$ and $b_1$ as follows:
\begin{eqnarray}
{\cal M}_{a_1}^2(y)\phi_{a_1}^{(i)}(y)&\equiv&
\Bigl[-\frac{d^2}{dy^2}+\frac{3}{2}
\abs{g\vev{A_1(y)}}^2-\Lambda^2\Bigr]
\phi_{a_1}^{(i)}(y)=(m_{a_1}^{(i)})^2\phi_{a_1}^{(i)}(y),\nonumber\\
{\cal M}_{b_1}^2(y)\phi_{b_1}^{(i)}(y)&\equiv&
\Bigl[-\frac{d^2}{dy^2}+\frac{1}{2}
\abs{g\vev{A_1(y)}}^2+\Lambda^2\Bigr]
\phi_{b_1}^{(i)}(y)=(m_{b_1}^{(i)})^2\phi_{b_1}^{(i)}(y).
\label{b2op}
\end{eqnarray}
\par
By noting
$\vev{A_1(y)}=\frac{2k\omega}{g}{\rm sn}
(\omega y, k)$ and by defining $u\equiv \omega y$,
we may recast the equation (\ref{a0boson}) into
\begin{equation}
\Bigl[-\frac{d^2}{du^2}+4k^2{\rm sn}^2(u, k)\Bigr]\phi_{a_0}^{(i)}(u)
=\Omega_{a_0}^{(i)}\phi_{a_0}^{(i)}(u),
\label{a0eigen}
\end{equation}
where
$\Omega_{a_0}^{(i)}\equiv \frac{(m_{a_0}^{(i)})^2}{\omega^2}$
with $\phi_{a_0}^{(i)}(u+2K(k))=\phi_{a_0}^{(i)}(u)$.
The same eigenvalue equation holds for $b_0$. For $a_1$ and $b_1$,
we also obtain
\begin{eqnarray}
\Bigl[-\frac{d^2}{du^2}+6k^2{\rm sn}^2(u, k)\Bigr]\phi_{a_1}^{(i)}(u)
&=&\Omega_{a_1}^{(i)}\phi_{a_1}^{(i)}(u),\\
\Bigl[-\frac{d^2}{du^2}+2k^2{\rm sn}^2(u, k)\Bigr]\phi_{b_1}^{(i)}(u)
&=&\Omega_{a_0}^{(i)}\phi_{b_1}^{(i)}(u),
\end{eqnarray}
where
$\Omega_{a_1}^{(i)}\equiv \frac{(m_{a_1}^{(i)})^2}{\omega^2}+1+k^2$
with $\phi_{a_1}^{(i)}(u+2K(k))=-\phi_{a_1}^{(i)}(u)$,
and
$\Omega_{b_1}^{(i)}\equiv \frac{(m_{b_1}^{(i)})^2}{\omega^2}-1-k^2$
with $\phi_{b_1}^{(i)}(u+2K(k))=-\phi_{b_1}^{(i)}(u)$.
Here we have used the relation $\omega = \Lambda/\sqrt{1+k^2}$.
\par
It immediately follows from the above analysis that
in order to find the bosonic mass spectrum for $R > R^*$
we have to solve the eigenvalue equation of the type
\begin{equation}
\Bigl[-\frac{d^2}{du^2}+N(N+1)k^2{\rm sn}^2(u, k)\Bigr]\phi^{(i)}(u)
=\Omega^{(i)}\phi^{(i)}(u)
\label{lameeq}
\end{equation}
with the boundary conditions $\phi^{(i)}(u+2K(k))=\pm \phi^{(i)}(u)$.
This equation is known as the Lam\'e equation.
The boson mass spectrum and its behavior with respect to the radius
$R$ are well understood by known properties of the Lam\'e equation,
though the exact eigenvalues and associated eigenfunctions are known
only for some special cases.
We summarize important properties of the eigenvalues and the
associated eigenfunctions
of the Lam\'e equation in the Appendix B.
\par
Before discussing the general case of $0\leq k < 1$, it may be
instructive to study the limit of
$k \rightarrow 0 ~(R\rightarrow R^*=1/{2\Lambda})$.
In this limit, the Lam\'e equation becomes
\begin{equation}
-\frac{d^2}{du^2}\phi(u, k=0)=\Omega(k=0)\phi(u,k=0).
\end{equation}
The boundary conditions we imposed are reduced to
$\phi(u+\pi)=\pm \phi(u)$ because of $2K(0)=\pi$.
It is easy to solve the equation. The eigenvalues and the
eigenfunctions at $k=0$ are simply given as in the Table III-3.
We can expand the fields $a_0$, $b_0$, $a_1$ and $b_1$
in terms of those
eigenfunctions as (\ref{a0expan1}) shown in the Appendix B.
\par
Now, let us come back to the original Lam\'e equation
with $0 \leq k <1$.
The solutions of the Lam\'e equation with the boundary conditions
$\phi(u+2K(k))=\pm\phi(u)$
are known to be classified by the four types of the eigenfunctions
\begin{equation}
Ec_N^{2n}(u,k),~Es_N^{2n+2}(u,k),~
~Ec_N^{2n+1}(u,k),~Es_N^{2n+1}(u,k)
\label{zeigen}
\end{equation}
with $n=0, 1, 2, \cdots$. The first two eigenfunctions in (\ref{zeigen})
are periodic under $u\rightarrow u+2K(k)$ and the last two ones are
antiperiodic. The first and the third eigenfunctions in
(\ref{zeigen})
are even under $u \rightarrow -u$ and the second and
the fourth ones are odd.
We denote the eigenvalues associated with $Ec_N^{n}$ and $Es_N^{n}$
by $\alpha_N^n(k)$ and  $\beta_N^n(k)$, respectively.
For a positive integer $N$, it has been known that
the lowest $2N+1$ eigenvalues and the associated eigenfunctions
are exactly known as Lam\'e polynomials, which are written in terms of
the Jacobi elliptic functions, ${\rm sn}(u, k), {\rm cn}(u, k)$
and ${\rm dn}(u, k)$. For general $N$, solutions of the Lam\'e
equation and even for integer $N$ other than $2N+1$ Lam\'e
polynomials will not be written in such the simple forms.
\par
Let us study the mass spectrum for the bosons
with the help of the known
analytic properties of the Lam\'e equation.
The Lam\'e equation for the field $a_1$
 corresponds to the $N=2$ case.
We may expand $a_1(x,y)$, taking
the boundary condition into account, as
\begin{equation}
a_1(x,y)=\sum_{n=1}^{\infty}\Bigl[
a_1^{(c,2n-1)}(x)Ec_2^{2n-1}(\omega y,k)
+a_1^{(s,2n-1)}(x)Es_2^{2n-1}(\omega y,k)
\Bigr].
\label{a1expan}
\end{equation}
The squared mass eigenvalues for
$a_1^{(c,2n-1)}$ and $a_1^{(s,2n-1)}$
are given by
\begin{equation}
(\alpha_2^{2n-1}(k)-1-k^2)\omega^2 \quad
{\rm and} \quad  (\beta_2^{2n-1}(k)-1-k^2)\omega^2,
\label{a1emode}
\end{equation}
respectively.
Two out of $2N+1 = 5$ Lam\'e polynomials satisfy the desired
antiperiodic boundary condition.
The two eigenfunctions $Ec_2^{1}(\omega y,k)$
and $Es_2^{1}(\omega y,k)$ are given by the Lam\'e
polynomials and their eigenvalues are exactly known.
Those eigenfunctions and eigenvalues are explicitly
given in the Table III-4.
The eigenvalues $\alpha_2^{2n-1}(k)$ and $\beta_2^{2n-1}(k)$
are degenerate at $k=0$, in fact, $\alpha_2^{2n-1}(0)=\beta_2^{2n-1}(0)
=(2n-1)^2$. They are still degenerate even for $k > 0$ except
for $\alpha_2^1(k)$ and $\beta_2^1(k)$.
By taking account of the mass hierarchies (\ref{hieboson})
in the Appendix B
and the degeneracy among the eigenvalues, the $R$-dependence
of the mass spectrum for $a_1^{(c,2n-1)}$ and $a_1^{(s,2n-1)}$
is schematically depicted in the Figure $1$.
\par
The Lam\'e equation for the field $b_1$
corresponds to the $N=1$ case.
We may expand the field as
\begin{equation}
b_1(x,y)=\sum_{n=1}^{\infty}\Bigl[
b_1^{(c,2n-1)}(x)Ec_1^{2n-1}(\omega y,k)
+b_1^{(s,2n-1)}(x)Es_1^{2n-1}(\omega y,k)
\Bigr].
\label{b1expan}
\end{equation}
Then, the squared mass eigenvalues for $b_1^{(c,2n-1)}$ and
$b_1^{(s,2n-1)}$ are given by
\begin{equation}
(\alpha_1^{2n-1}(k)+1+k^2)\omega^2 \quad
{\rm and} \quad (\beta_1^{2n-1}(k)+1+k^2)\omega^2,
\label{b1emode}
\end{equation}
respectively.
Two out of $2N+1 = 3$ Lam\'e polynomials satisfy the desired
antiperiodic boundary condition.
The two eigenfunctions $Ec_1^{1}(\omega y,k)$
and $Es_1^{1}(\omega y,k)$ are given by the Lam\'e
polynomials and their eigenvalues are exactly known.
Those eigenfunctions and eigenvalues are explicitly
given in the Table III-5.
The eigenvalues $\alpha_1^{2n-1}$ and $\beta_1^{2n-1}$ are degenerate at
$k=0$, and they are still degenerate even for $k>0$ except for
$\alpha_1^1(k)$ and $\beta_1^1(k)$. The $R$-dependence
of the mass spectrum
for $b_1^{(c,2n-1)}$ and $b_1^{(s,2n-1)}$ is schematically depicted
in the Figure $1$.
\par
Although the $2N+1$ Lam\'e polynomials for integer $N$ are
known exactly, those exact analytical solutions to the Lam\'e
equation are rather exceptional. The Lam\'e equation for the
fields $a_0$ and $b_0$ corresponds to noninteger $N$, and
none of the eigenfunctions and the eigenvalues are known exactly.
Even though we do not know the exact solutions
for this case, we can study the qualitative behaviour of the mass
spectrum of $a_0$ and $b_0$.
We can expand the fields $a_0$ and $b_0$ as
\begin{eqnarray}
a_0(x,y)&=&\sum_{n=0}^{\infty}\Bigl[
a_0^{(c,2n)}(x)Ec_{N_0}^{2n}(\omega y,k)+
a_0^{(s,2n+2)}Es_{N_0}^{2n+2}(\omega y,k)\Bigr], \nonumber\\
b_0(x,y)&=&\sum_{n=0}^{\infty}\Bigl[
b_0^{(c,2n)}(x)Ec_{N_0}^{2n}(\omega y,k)+
b_0^{(s,2n+2)}Es_{N_0}^{2n+2}(\omega y,k)
\Bigr],
\label{a0expan}
\end{eqnarray}
where $N_0$ satisfies $N_0(N_0+1)=4$.
The squared mass eigenvalue for
each mode $a_0^{(c,2n)}$, $a_0^{(s,2n+2)}$,
$b_0^{(c,2n)}$ and $b_0^{(s,2n+2)}$
is given by
\begin{eqnarray}
a_0^{(c,2n)}, b_0^{(c,2n)};& &
\alpha_{N_0}^{2n}(k)\omega^2, \nonumber\\
a_0^{(s,2n+2)}, b_0^{(s,2n+2)};& &\beta_{N_0}^{2n+2}(k)\omega^2.
\label{a0emode}
\end{eqnarray}
Since we have already known
$\alpha_{N_0}^{2n}(k=0)=\beta_{N_0}^{2n}(k=0)
=4n^2$, the modes $a_0^{(c,0)}$ and $b_0^{(c,0)}$ are massless at $k=0$.
They are, however, expected to become massive for $k>0$. This is
concluded from the eigenvalue equation (\ref{a0eigen}).
The differential operator
${\hat H}\equiv -\frac{d^2}{du^2}+4k^2{\rm sn}^2(u,k)$
is positive definite for $k>0$, so that
$\vev{\phi |{\hat H}|\phi}>0$ for any nontrivial
eigenfunctions. Therefore, we
conclude $\alpha_{N_0}^0(k)$, $\beta_{N_0}^0(k) > 0$ for $k>0$.
The eigenvalues $\alpha_{N_0}^{2n}$ and $\beta_{N_0}^{2n}$ are
degenerate at $k=0$, but they split for $k>0$ though we do not know
their
relative magnitude. The $R$-dependence of the mass eigenvalues of
$a_0^{(c,2n)}$, $a_0^{(s,2n+2)}$, $b_0^{(c,2n)}$ and $b_0^{(s,2n+2)}$
is schematically depicted in the Figure $1$.
\par
Since we have studied the boson mass
spectrum at the critical radius $R=R^* (k=0)$, it may be instructive
to study the behaviour of the mass spectrum for small values of $k$
in perturbation theory with respect to
the parameter $k$.
In the Tables III-6 and 7, the perturbative mass spectrum
for the first two lowest modes of $a_0$, $b_0$ and for the
second excited modes of $a_1$, $b_1$ are summarized,
up to the order of $k^2$.
\subsubsection{Fermion sector}
We shall compute the mass spectrum of the fermions.
The relevant part of the Lagrangian is
\begin{equation}
{\cal L}_{F(R>R^*)}^{(2)}=\int_0^{2\pi R}dy\Bigl[
-i{\bar\psi}_0{\bar{\sigma}}^M\del_M\psi_0
-i{\bar\psi}_1{\bar{\sigma}}^M\del_M\psi_1
+(g\vev{A_1(y)}\psi_0\psi_1+h.c.)
\Bigr].
\label{zrfermi}
\end{equation}
As before, according to our prescription (\ref{diracs}) given
in the Appendix B, we rewrite (\ref{zrfermi}) in terms
of the 3-dimensional
Majorana spinors (\ref{majo}) as follows:
\begin{equation}
{\cal L}_{F(R>R^*)}^{(2)}=\int_0^{2\pi R}dy~
\frac{1}{2}{\bar\Psi}(-i\gamma^{\mu}\del_{\mu}{\bf 1}
+{\cal M}_F(y))\Psi,
\end{equation}
where we have defined $\Psi\equiv (\chi_0, \rho_0, \chi_1, \rho_1)^T$
and
\begin{equation}
{\cal M}_F(y)\equiv\left(\begin{array}{cccc}
0&-\del_y&U(y)&0\\
\del_y&0&0&-U(y)\\
U(y)&0&0&-\del_y\\
0&-U(y)&\del_y&0
\end{array}\right)
\end{equation}
where
$U(y)\equiv 2k\omega~{\rm sn}(\omega y, k)$.
The field equations for $\chi_0$, $\rho_0$, $\chi_1$ and $\rho_1$
are easily obtained. Multiplying the field equations by
$(i\gamma^{\nu}\del_{\nu}{\bf 1}+{\cal M}_F)$,
we have the second order differential equations for these fields
\begin{equation}
(-\del^{\mu}\del_{\mu}{\bf 1}+{\cal M}_F^2)\Psi=
\left(\begin{array}{cccc}
-\del^{\mu}\del_{\mu}+D_y^2&0&0&\del_yU(y)\\
0&-\del^{\mu}\del_{\mu}+D_y^2&\del_yU(y)&0\\
0&\del_yU(y)&-\del^{\mu}\del_{\mu}+D_y^2&0\\
\del_yU(y)&0&0&-\del^{\mu}\del_{\mu}+D_y^2
\end{array}\right)\Psi=0.
\label{fermiop}
\end{equation}
Here, we have defined $D_y^2\equiv -\del_y^2+(U(y))^2$.
It is convenient to shuffle $\chi_0$, $\rho_0$, $\chi_1$
and $\rho_1$ as follows:
\begin{equation}
\eta_{\pm}\equiv \frac{1}{\sqrt 2}(\chi_0\pm \rho_1),\quad
\zeta_{\pm}\equiv \frac{1}{\sqrt 2}(\rho_0\pm \chi_1).
\end{equation}
Then, the Majorana spinors $\eta_{\pm}$
and $\zeta_{\pm}$ satisfy the equations of motion
\begin{eqnarray}
\bigl[-\del^{\mu}\del_{\mu}
-\del_y^2+(U(y))^2 \pm \del_y U(y)\bigr]\eta_{\pm}(x^{\mu},y)
&=& 0, \nonumber\\
\bigl[-\del^{\mu}\del_{\mu}
-\del_y^2+(U(y))^2 \pm \del_y U(y)\bigr]\zeta_{\pm}(x^{\mu},y)
&=& 0.
\label{nnewfeq}
\end{eqnarray}
To find the mass spectrum, we have to be careful about the boundary
conditions. The boundary conditions for $\eta_{\pm}$, $\zeta_{\pm}$
are translated into
\begin{equation}
\eta_{\pm}(x^{\mu}, y+2\pi R)=\eta_{\mp}(x^{\mu},y),\qquad
\zeta_{\pm}(x^{\mu}, y+2\pi R)=\zeta_{\mp}(x^{\mu},y),
\label{newbcf}
\end{equation}
which are also equivalent to
\begin{eqnarray}
\eta_{+}(x^{\mu}, y+4\pi R)&=&\eta_{+}(x^{\mu},y),\qquad
\eta_{-}(x^{\mu},y)=\eta_{+}(x^{\mu},y+2\pi R),\nonumber\\
\zeta_{+}(x^{\mu}, y+4\pi R)&=&\zeta_{+}(x^{\mu},y), \qquad
\zeta_{-}(x^{\mu},y)=\zeta_{+}(x^{\mu},y+2\pi R).
\label{newrel}
\end{eqnarray}
Note that these are consistent with the fields equations because of
$U(y+2\pi R)=-U(y)$.
Once we solve the equations of motion for
$\eta_{+}$, $\zeta_{+}$ with the
boundary conditions (\ref{newrel}), we automatically have solutions
for $\eta_{-}, \zeta_{-}$ through the relations (\ref{newrel}).
It follows from the above consideration
that the mass spectrum of the
fermions is obtained by solving the eigenvalue equation
\begin{equation}
\Bigl[-\frac{d^2}{dy^2}+(U(y))^2+\frac{dU(y)}{dy}\Bigr]
\phi(y)=m_F^2\phi(y)
\label{fermieigen}
\end{equation}
with the boundary condition $\phi(y+4\pi R)=\phi(y)$.
Changing the variables $y$ to $u\equiv \omega y$,
we recast the above equation into
\begin{equation}
\Bigl(\frac{d}{du}+2k~{\rm sn}(u, k)\Bigr)
\Bigl(-\frac{d}{du}+2k~{\rm sn}(u, k)\Bigr)
\phi(u)=(\frac{m_F}{\omega})^2\phi(u)
\label{heihou}
\end{equation}
with $\phi(u+4K(k))=\phi(u)$.
\par
Let us first study the behavior of the fermion mass
spectrum in the limit of $k\rightarrow 0$
(or $R \rightarrow R^*=1/{2\Lambda}$).
In this limit the above equation is reduced to
\begin{equation}
-\frac{d^2}{du^2}\phi(u,k=0)=
(\frac{m_F(k=0)}{\omega})^2\phi(u,k=0)
\end{equation}
with $\phi(u+2\pi, k=0)=\phi(u,k=0)$.
The eigenvalues and the associated eigenfunctions
are easily obtained as in the Table III-8.
\par
The properties of the eigenvalues and the eigenfunctions of
the equation (\ref{fermieigen}) is summarized in the Appendix B.
As discussed there, in terms of
the eigenfunctions denoted by
$Ec^{\prime n}(u,k)$ and $Es^{\prime n+1}(u,k)~(n=0,1, 2\cdots)$,
we can expand
$\eta_{+}(x^{\mu}, y)$ and $\zeta_{+}(x^{\mu}, y)$ as
\begin{eqnarray}
\eta_{+}(x^{\mu}, y)&=&\sum_{n=0}^{\infty}\Bigl[
\eta^{(c,n)}(x)Ec^{\prime n}(\omega y, k)+
\eta^{(s,n+1)}(x)Es^{\prime n+1}(\omega y, k)\Bigr],\nonumber\\
\zeta_{+}(x^{\mu}, y)&=&\sum_{n=0}^{\infty}\Bigl[
\zeta^{(c,n)}(x)Ec^{\prime n}(\omega y, k)+
\zeta^{(s,n+1)}(x)Es^{\prime n+1}(\omega y, k)\Bigr].
\label{fexp1}
\end{eqnarray}
Then, through the relations (\ref{newrel}), $\eta_{-}(x^{\mu}, y)$
and $\zeta_{-}(x^{\mu}, y)$ are given by
\begin{eqnarray}
\!\!\!\!\!\!\eta_{-}(x^{\mu}, y)\!\!&=&\!\!\eta_{+}(x,y+2\pi
R)\nonumber\\
\!\!&=&\!\!\sum_{n=0}^{\infty}\Bigl[
\eta^{(c,n)}(x)Ec^{\prime n}(\omega (y+2\pi R), k)+
\eta^{(s,n+1)}(x)Es^{\prime n+1}(\omega(y+2\pi R), k)\Bigr],\nonumber\\
\!\!\!\!\!\!\zeta_{-}(x^{\mu}, y)\!\!&=&\!\!\zeta_{+}(x,y+2\pi R)
\nonumber\\
\!\!&=&\!\!\sum_{n=0}^{\infty}\Bigl[
\zeta^{(c,n)}(x)Ec^{\prime n}(\omega (y+2\pi R), k)+
\zeta^{(s,n+1)}(x)Es^{\prime n+1}(\omega (y+2\pi R), k)\Bigr].
\label{fexp2}
\end{eqnarray}
Equipped with these, it is easy to obtain
the equations of motion for $\eta^{(c,n)}(x)$,
$\eta^{(s,n)}(x)$, $\zeta^{(c,n)}(x)$ and $\zeta^{(s,n)}(x)$
from the field equations (\ref{nnewfeq}) as follows:
\begin{equation}
\Bigl[-\del_{\mu}\del^{\mu}+(m_F^{(i, n)})^2\Bigr]
\eta^{(i,n)}(x)=0,\quad
\Bigl[-\del_{\mu}\del^{\mu}+(m_F^{(i, n)})^2\Bigr]
\zeta^{(i,n)}(x)=0,
\label{newfeq1}
\end{equation}
where $i=c,s$ and we have denoted the eigenvalues of
$Ec^{\prime n}(u,k)$ and $Es^{\prime n}(u,k)$ by
$(m_F^{(c,n)}/\omega)^2$
and $(m_F^{(s,n)}/\omega)^2$, respectively.
\par
Even though we do not know exact results of the eigenvalue
equation (\ref{fermieigen}), we can extract
the behaviour of the fermion mass spectrum with
respect to the radius $R$ by applying general properties of the
equation. As proven in the Appendix B,
we have the hierarchies among the eigenvalues
of the eigenvalue equation.
They yield the mass hierarchies among the masses of the fermions
\begin{eqnarray}
m_F^{(c,0)} &<& m_F^{(c,1)} < m_F^{(c,2)} < \cdots, \nonumber\\
m_F^{(s,1)} &<& m_F^{(s,2)} < m_F^{(s,3)} < \cdots, \nonumber\\
m_F^{(c,n)}&=& m_F^{(s,n)},\qquad {\rm if}\quad n\neq 0.
\label{hiefermi}
\end{eqnarray}
According to these, we schematically depict the $R$-dependence of the
fermion mass spectrum for
$\eta^{(c,n)}$, $\eta^{(s,n)}$, $\zeta^{(c,n)}$ and $\zeta^{(s,n)}$
for $R>R^*$ in the Figure $1$.
\par
We do not know the exact eigenvalues of $(m_F^{(c, n)})^2$
and $(m_F^{(s, n)})^2$ except for the lowest eigenvalue.
Let us recast the eigenvalue
equation (\ref{heihou}) into the form
$\hat{a}^{\dagger}(u)\hat{a}(u)\phi(u)
= (\frac{m_F}{\omega})^2 \phi(u)$,
where
$\hat{a}(u) \equiv
-\frac{d}{du}+2k{\rm sn}(u,k)$ and
$\hat{a}^{\dagger}(u) \equiv
+\frac{d}{du}+2k{\rm sn}(u,k)$.
It immediately follows that all eigenvalues are positive semi-definite.
In fact, the lowest eigenvalue is found to be zero provided that the
eigenfunction satisfies
$\hat{a}(u)Ec^{\prime 0}(u,k) = 0$  with
$Ec^{\prime 0}(u+4K(k),k)=Ec^{\prime 0}(u,k)$.
The solution to the above equation is given, apart from a normalization
constant, by
$Ec^{\prime 0}(u,k)=
(k{\rm cn}(u,k)-{\rm dn}(u,k) )^2$.
This is consistent with the previous result at $k = 0$.
We can see from (\ref{fexp1}) that the modes $\eta^{(c,0)}(x)$
and $\zeta^{(c,0)}(x)$ are massless.
All the other modes are massive. It is not difficult to
see that these massless modes correspond to the Nambu-Goldstone fermions
associated with the spontaneous SUSY breaking, as we will discuss
in the next subsection.
\par
Although we do not know the explicit
forms of the eigenfunctions and their eigenvalues
except for the lowest mode,
we can obtain the perturbative mass spectrum for higher
modes with respect to $k$. The results for the first
three excited states are given in the Table III-9.
\par
\subsection{Analysis of the mass spectrum for $R>R^*$}
Let us discuss the boson and fermion mass spectra
obtained in the previous subsection 3.4.
We have found that the lowest mode of $a_1$ is  massless.
This mode is interpreted as the Nambu-Goldstone
boson associated with the breakdown of the translational invariance
for the $S^1$ direction.
To see this, let us recall the expression (\ref{a1expan}) for
$a_1(x,y)$ in terms of the eigenfunctions
 of the Lam\'e equation with $N=2$.
The Nambu-Goldstone mode has to
be the one proportional to $\del_y \vev{A_1(x,y)}$
in the expansion of $A_1(x,y)$ because of the commutation relation
$[ia Q, A_1(x,y)]=a\del_y A_1(x,y)$, where $Q$ is a generator of the
translation for the $S^1$ direction.
Noting that
$Ec_2^{1}(\omega y,k) = {\rm cn}(\omega y,k){\rm dn}(\omega y,k)
\propto \frac{d}{dy}\langle A_1(x,y) \rangle$,
we find that the coefficient of $Ec_2^{1}(\omega y,k)$, {\it i.e.}
$a_1^{(c,1)}(x)$ corresponds to the Nambu-Goldstone boson associated
with the spontaneous breakdown of the translational invariance for the
$S^1$ direction.
Originally, the mode $d\vev{A_1}/dy$ is a local tangent specified by
one parameter, which is a consequence of the translational invariance
of the equation of motion, to the space given by all the translated
solutions $\vev{A_1(y-a)},~a\in {\bf R}$. So, there always exists
such the zero mode on the quantization around the coordinate-dependent
background.
\par
It is important to notice that the translational invariance of $S^1$ can
be
interpreted as a global $U(1)$ invariance from the $3$-dimensional point
of
view. In this point of view, a massless bosonic mode in the phase of the
spontaneous breakdown of the translational invariance can be interpreted
as an ordinary Nambu-Goldstone boson associated
with the spontaneous breakdown of the
global $U(1)$ symmetry. To see this, let us expand $A_i (i=0,1)$ in
Fourier series, according to the boundary conditions.
The commutation relations between the generator $Q$ and the Fourier
modes may be given by
\begin{eqnarray}
\left[Q,~{\tilde A}_0^{(n)}(x)\right]
&=&\frac{n}{R}~{\tilde A}_0^{(n)}(x), \quad n \in{\bf Z} , \nonumber\\
\left[Q,~{\tilde A}_1^{(l)}(x)\right]
&=&\frac{l}{R}~{\tilde A}_1^{(l)}(x),
\quad l \in {\bf Z}+\half .
\end{eqnarray}
We see that the generator $Q$, which is
originally the generator of the
translation of $S^1$, acts as if it generates global
$U(1)$ transformations under which the fields ${\tilde A}_0^{(n)}$
and ${\tilde A}_1^{(l)}$
carry the $U(1)$ charges $n/R$ and $ l/R$, respectively.
For $R\leq R^*$,
the vacuum expectation
values for $A_{0}$ and $A_{1}$ are
an arbitrary constant and zero, respectively, so that
$\vev{{\tilde A}_0^{(0)}}={\rm arbitrary~constant}$,
$\vev{{\tilde A}_0^{(n \neq 0)}}=0$ and $\vev{{\tilde A}_1^{(l)}}=0$.
Since ${\tilde A}_0^{(0)}$ carries no $U(1)$ charge,
the nonvanishing vacuum expectation
value of ${\tilde A}_0^{(0)}$ does not break the $U(1)$ symmetry.
For $R>R^*$, the vacuum
expectation values are given by (\ref{zvac2}).
Since ${\rm sn}(\omega y,k)$ can be expanded in Fourier series as
${\rm sn}(\omega(y-y_0),k)=\sum_{l\in{\bf Z}+\half >0}^{\infty}
c_l\sin (l\frac{(y-y_0)}{R})$ with $c_l\neq 0$,
we obtain $\vev{{\tilde A}_0^{(0)}}=\frac{\mu}{g}\sqrt{2\pi R},~
\vev{{\tilde A}_0^{(n\neq 0)}}=0$ and $\vev{{\tilde A}_1^{(l)}}\neq 0$.
The ${\tilde A}_1^{(l)}$ has a $U(1)$ charge $l/R$,
so that the vacuum expectation
value $\vev{\tilde A_1^{(l)}}$ breaks the $U(1)$
symmetry spontaneously. This observation is
consistent with the previous discussion that the Nambu-Goldstone
boson is incorporated in $A_1(x,y)$ as $\del_y \vev{A_1(x,y)}$.
\par
Let us discuss the Nambu-Goldstone fermions associated with the
breakdown of the supersymmetry. Even though we do not obtain the
fermion mass spectrum exactly, we
definitely have massless modes associated with the eigenfunction
$Ec^{\prime 0}(u,k)$. The vacuum expectation values of the
SUSY transformations for the spinors
in the vacuum configuration (\ref{zvac2}) are given by
\begin{eqnarray}
\vev{\delta_S\psi_{0\alpha}(x,y)}&=&-\sqrt{2}\xi_{\alpha}
(\frac{\Lambda^2}{g}-\frac{g}{2}\langle A_1^*(y)\rangle^2),
\nonumber\\
\vev{\delta_S\psi_{1\alpha}(x,y)}&=&-\sqrt{2}(\xi_{\alpha})^*
\frac{d\vev{A_1(y)}}{dy},
\label{sutra2}
\end{eqnarray}
where we have used ${\bar\xi}^{\dot\beta}=(i\tau_y\xi^*)_{\beta}$
and $\sigma^y=\tau_y$.
Since both $\vev{\delta_S\psi_{0}}$ and
$\vev{\delta_S\psi_{1}}$ do not vanish for any nontrivial $\xi$,
the supersymmetry is completely broken spontaneously and no
SUSY is left in $3$-dimensions.
In the Appendix B we present the expressions for the expansion of
$\psi_0$ and $\psi_1$ in terms of the eigenfunctions
$Ec^{\prime n}$ and $Es^{\prime n}$
(see (\ref{ngexp1}) and (\ref{ngexp2})).
Since $Ec^{\prime 0}$ is found to be
$Ec^{\prime 0}=(k~{\rm cn}(u,k)-{\rm dn}(u,k))^2$, we can show that
\begin{eqnarray}
\psi_0(x,y)&=&\frac{g}{\omega^2}(\eta^{(c,0)}(x)+i\zeta^{(c,0)}(x))
(\frac{\Lambda^2}{g}-\frac{g}{2}\vev{A_1^*(y)}^2)+\cdots ,\nonumber\\
\psi_1(x,y)&=&\frac{g}{\omega^2}(\eta^{(c,0)}(x)+i\zeta^{(c,0)}(x))^*
\frac{d\vev{A_1(y)}}{dy}+\cdots,
\label{fzerom}
\end{eqnarray}
where $\cdots$ denotes higher modes.
Comparing (\ref{fzerom}) with (\ref{sutra2})
and identifying $\xi$  in (\ref{sutra2}) with
$\frac{g}{\sqrt 2\omega^2}(\eta^{(c,0)}+i\zeta^{(c,0)})$,
we can conclude
that the fermion zero modes $\eta^{(c,0)}$ and $\zeta^{(c,0)}$
are nothing but the Nambu-Goldstone fermions associated with the
spontaneous SUSY breaking.
\par
Let us comment on the supertrace formula for the mass spectrum
in the $Z_2$-model for $R>R^*$.
We can immediately see
from (\ref{b1op}), (\ref{b2op}) and (\ref{fermiop})
that the supertrace formula
${\rm Tr}~{\cal M}_B^2={\rm Tr}~{\cal M}_F^2$
formally holds.
This relation, however, may mathematically be meaningless because the
trace has to be taken in an infinite dimensional space, so that the
sum of the eigenvalues of each squared mass operator would diverge.
For $R\le R^*$, the relations (\ref{sutramass}) hold in the subspace
of each Fourier mode. So, they are meaningful relations. We do not
know whether there are any finite dimensional
subspaces in which the relation still holds for $R>R^*$.
Finally, we summarize the phase structure of the model in
the Table III-10.
\section{THE ${\bf U(1)}$ MODEL}
In this section we shall study the $U(1)$ model which
has a global $U(1)$ symmetry, instead of the discrete $Z_2$-symmetry.
We will find interesting features different from those of the
$Z_2$ model. We, again, study the vacuum structure and the mass
spectrum for bosons and fermions.
\subsection{Vacuum configuration}
The superpotential of the $U(1)$ model is given by
\begin{equation}
W(\Phi_0,\Phi_{\pm}) = g\Phi_0 \bigg(\frac{\Lambda^2}{g^2}
-\Phi_+ \Phi_- \bigg).
\label{u1w}
\end{equation}
We shall use this superpotential which has no mass term for
$\Phi_{\pm}$.
We have slightly changed the notations of the superpotential
(\ref{equp}).
The model has two global $U(1)$ and $U(1)_R$
symmetries. The $U(1)$ symmetry is defined by
$\Phi_0 \longrightarrow +\Phi_0$ and
$\Phi_{\pm} \longrightarrow e^{\pm2\pi i\alpha}\Phi_{\pm}$.
This global symmetry plays an important role for our mechanism to work.
The scalar potential $V(A_0,A_{\pm})$ is given by
\begin{eqnarray}
V(A_0,A_{\pm}) &=&\abs{F_0}^2+\abs{F_+}^2+\abs{F_{-}}^2 \nonumber \\
&=& \bigg|\frac{\Lambda^2}{g}-gA_+ A_{-} \bigg|^2 +
|gA_0|^2\bigg(|A_+|^2+|A_-|^2 \bigg).
\end{eqnarray}
The solutions to the $F$-term conditions are easily found
to be $\bar{A_0}=0$ and $\bar{A}_+\bar{A}_-=\frac{\Lambda^2}{g^2}$,
at which the scalar potential vanishes.
\par
Let us consider the model on $M^3$ $\otimes$ $S^1$.
The notations concerning about
the space-time are the same with the previous section.
Using the $U(1)$ symmetry degrees of freedom,
we can impose the nontrivial boundary
conditions on the superfields $\Phi_0(x^{\mu},y)$ and
$\Phi_{\pm}(x^{\mu},y)$ for the $S^1$ direction
\begin{equation}
\Phi_0(x^{\mu},y+2\pi R) = +\Phi_0(x^{\mu},y),\quad
\Phi_{\pm}(x^{\mu},y+2\pi R)
= e^{\pm 2\pi i\alpha}\Phi_{\pm}(x^{\mu},y).
\label{u1bc}
\end{equation}
It should be emphasized that
we do not use the $U(1)_R$
symmetry degrees of freedom to impose nontrivial boundary
conditions, as stated earlier.
In the following we restrict the phase $\alpha$
to be the range $0 < \alpha \leq \frac{1}{2}$ without loss of
generality.
\par
It is easy to see that the $F$-term conditions are not consistent with
the boundary conditions (\ref{u1bc}),
by which the vacuum expectation values of
$A_{\pm}$ are forced not to take any nonzero constants. Therefore, the
vacuum configuration given by the solutions of the $F$-term conditions
are not realized as a supersymmetric vacuum configuration.
Thus, our mechanism works again in this model.
\par
As we have done in the previous section, in order
to find the true vacuum
configuration, we have to solve the minimization
problem of the ``energy'' functional defined by
\begin{eqnarray}
{\cal E}[A_0,A_{\pm};R] &\equiv&
\int_0^{2\pi R}dy \bigg[ \bigg|\frac{dA_0}{dy}\bigg|^2+
\bigg|\frac{dA_+}{dy}\bigg|^2+\bigg|\frac{dA_-}{dy}\bigg|^2 \nonumber\\
& & + \bigg|\frac{\Lambda^2}{g}-gA_+ A_- \bigg|^2 +
|gA_0|^2\bigg(|A_+|^2+|A_-|^2 \bigg) \bigg].
\label{u1efunc}
\end{eqnarray}
Then, the vacuum configuration for $A_0$ and $A_{\pm}$ should satisfy
the field equations
\begin{eqnarray}
0 &=& \frac{\delta{\cal E}[A_0,A_{\pm};R]}{\delta A_0^*(y)}
= -\frac{d^2 A_0(y)}{dy^2}+g^2A_0(y)\bigg(
|A_+(y)|^2+|A_-(y)|^2\bigg), \\ \nonumber
0 &=& \frac{\delta{\cal E}[A_0,A_{\pm};R]}{\delta A_{\pm}^*(y)}
\nonumber \\
&=& -\frac{d^2 A_{\pm}(y)}{dy^2}-g^* A_{\mp}^*(y)\bigg(
\frac{\Lambda^2}{g}-gA_+(y)A_-(y) \bigg)
+g^2 |A_0(y)|^2 A_{\pm}(y).
\end{eqnarray}
It follows from the first equation that the vacuum configuration
can be classified into two types of solutions as
\begin{equation}
({\rm I});~\left\{\begin{array}{ll}
A_0(y)=& {\rm arbitrary~constant}\\
A_{\pm}(y)=& 0\\
\end{array}\right.
~~({\rm II});~\left\{\begin{array}{ll}
A_0(y)=& 0\\
A_{+}(y)\neq & 0 ~~{\rm and/or}~~ A_{-}\neq 0. \\
\end{array}\right.
\end{equation}
For the type (I) solution,
${\cal E}[A_0 , A_{\pm};R]$ becomes
\begin{equation}
{\cal E}[A_0 = {\rm const.}, A_{\pm}=0;R] = \frac{2\pi R
\Lambda^4}{g^2}.
\label{etype1}
\end{equation}
On the other hand, for the type (II) solution,
we can easily show that
\begin{eqnarray}
{\cal E}[A_0 =0, A_{\pm};R]\bigg|
_{\frac{\delta{\cal E}}{\delta A_i}=0}
&=& \frac{2\pi R \Lambda^4}{g^2}
-g^2\int_0^{2\pi R}dy|A_+(y)A_-(y)|^2, \nonumber \\
&\leq& {\cal E}[A_0 = {\rm const.}, A_{\pm}=0;R].
\label{type2}
\end{eqnarray}
An important conclusion here is that if there appears any
type (II) solution with $A_{\pm} \neq 0$,
then, the type (I) solution
is no longer a vacuum configuration.
\par
Let us next find the vacuum configuration explicitly. To do
this, we redefine the fields
\begin{equation}
A_0(y) \equiv \tilde{A}_0(y), \qquad
A_{\pm}(y) \equiv e^{\pm i\frac{\alpha}{R}y}
\tilde{A}_{\pm}(y).
\label{u1redef}
\end{equation}
Note that
$\tilde{A}_0(y)$ and $\tilde{A}_{\pm}(y)$ satisfy the
periodic boundary condition. Inserting
(\ref{u1redef}) into (\ref{u1efunc}), we obtain
\begin{equation}
{\cal E}[A_0,A_{\pm};R] = {\cal E}_{KE}[\tilde{A}_0;R]+
{\cal E}_{KE}[\tilde{A}_+;R]+{\cal E}_{KE}[\tilde{A}_-;R]
+{\cal E}_{PE}[\tilde{A}_0,\tilde{A}_{\pm};R],
\label{u1func2}
\end{equation}
where
\begin{eqnarray}
\!\!\!\!\!\!{\cal E}_{KE}[\tilde{A}_0;R] \!\!&=& \!\!\int_0^{2\pi
R}\!\!dy
\bigg|\frac{d\tilde{A}_0}{dy} \bigg|^2, \nonumber\\
\!\!\!\!\!\!{\cal E}_{KE}[\tilde{A}_{\pm};R] \!\!&=& \!\!\int_0^{2\pi
R}\!\!dy
\bigg[\bigg|\frac{d\tilde{A}_{\pm}}{dy} \bigg|^2
\mp i\frac{\alpha}{R}\bigg(\tilde{A}_{\pm}^*
\frac{d\tilde{A}_{\pm}}{dy}-\frac{d\tilde{A}_{\pm}^*}{dy}
\tilde{A}_{\pm} \bigg) \bigg], \nonumber\\
\!\!\!\!\!\!{\cal E}_{PE}[\tilde{A}_0,\tilde{A}_{\pm};R]
\!\!&=& \!\!\int_0^{2\pi R}\!\!dy
\bigg[ \bigg|\frac{\Lambda^2}{g}-g\tilde{A}_+\tilde{A}_- \bigg|^2
+ \bigg(g^2|\tilde{A}_0|^2+\bigg(\frac{\alpha}{R}\bigg)^2 \bigg)
(|\tilde{A}_+|^2+|\tilde{A}_-|^2)\bigg].
\label{u1func3}
\end{eqnarray}
Our strategy to find the vacuum configuration is that we first look
for configurations which minimize each of
${\cal E}_{KE}[\tilde{A}_0;R]$, ${\cal E}_{KE}[\tilde{A}_{\pm};R]$
and ${\cal E}_{PE}[\tilde{A}_0,\tilde{A}_{\pm};R]$ and then construct
configurations which minimize all of them.
\par
Let us look for configurations which minimize each of
${\cal E}_{KE}[\tilde{A}_0;R]$ and ${\cal E}_{KE}[\tilde{A}_{\pm};R]$.
For that purpose, we expand the fields in Fourier series,
taking account of the boundary conditions, as
$\tilde{A}_0(y) = \sum_{n\in{\bf Z}}a_0^{(n)}e^{i\frac{n}{R}y}$ and
$\tilde{A}_{\pm}(y) = \sum_{n\in{\bf Z}}a_{\pm}^{(n)}e^{i\frac{n}{R}y}$.
Inserting them into
${\cal E}_{KE}[\tilde{A}_0;R]$ and ${\cal E}_{KE}[\tilde{A}_{\pm};R]$,
we obtain
\begin{equation}
{\cal E}_{KE}[\tilde{A}_0;R] =
\frac{2\pi}{R}\sum_{n \in {\bf Z}}n^2|a_0^{(n)}|^2, \quad
{\cal E}_{KE}[\tilde{A}_{\pm};R] =\frac{2\pi}{R}
\sum_{n \in {\bf Z}}
[(n \pm \alpha)^2-\alpha^2]|a_{\pm}^{(n)}|^2.
\end{equation}
Since $\alpha$ is restricted to the range
$0 < \alpha \leq \frac{1}{2}$, ${\cal E}_{KE}[\tilde{A}_0;R]$
and ${\cal E}_{KE}[\tilde{A}_{\pm};R]$ are positive semi-definite.
The configurations which give
${\cal E}_{KE}[\tilde{A}_0;R]=0$ and
${\cal E}_{KE}[\tilde{A}_{\pm};R]=0$ are found to be
\begin{eqnarray}
{\rm for}~~0<\alpha <\half, ~~& &
\left\{\begin{array}{l}
{\tilde A}_0(y)= a_0^{(0)},\\
{\tilde A}_{\pm}(y)= a_{\pm}^{(0)},\\
\end{array}\right.\\
 {\rm for}~~\alpha=\half,~~ & &\left\{\begin{array}{l}
{\tilde A}_0(y)=  a_0^{(0)},\\
{\tilde A}_{\pm}(y)=a_{\pm}^{(0)}+a_{\pm}^{(\mp 1)}\e^{\mp iy/R},
\end{array}\right.
\end{eqnarray}
where $a_0^{(0)}$, $a_{\pm}^{(0)}$ and $a_0{\pm}^{(\mp 1)}$
are arbitrary complex constants.
Let us next look for configurations which minimize
${\cal E}_{PE}[\tilde{A}_0,\tilde{A}_{\pm};R]$. As we saw before,
we have the two types of the solutions, {\it i.e.} type (I) and (II).
For the case (I), ${\cal E}_{PE}[\tilde{A}_0,\tilde{A}_{\pm};R]$
becomes (\ref{etype1}). To find the type (II) solution
we rewrite $\tilde{A}_{\pm}$ into the form
$\tilde{A}_{+}(y) = \rho_+(y)e^{i\theta(y)}$ and
$\tilde{A}_{-}(y) = \rho_-(y)e^{-i\theta(y)+i\phi(y)}$,
where $\rho_{\pm}(y)$ are real and positive semi-definite
functions of $y$ and $\theta(y)$, $\phi(y)$ are real functions
of $y$. Inserting them into ${\cal E}_{PE}$, we obtain
\begin{eqnarray}
{\cal E}_{PE}[\tilde{A}_0=0,\tilde{A}_{\pm};R] &\equiv&
{\cal E}_{PE}[\rho_{\pm},\theta, \phi ;R] \nonumber \\
&=& \int_0^{2\pi R}dy\bigg[\bigg|\frac{\Lambda^2}{g}
-g \rho_+ \rho_- e^{i\phi}\bigg|^2 +
\bigg(\frac{\alpha}{R}\bigg)^2
\bigg((\rho_+)^2+(\rho_-)^2 \bigg)\bigg].
\end{eqnarray}
Since $\rho_+$ and $\rho_-$ are positive semi-definite and since
$\rho_+\rho_- \neq 0$ for the configurations which minimize
${\cal E}_{PE}[\rho_{\pm},\theta, \phi ;R]$,
$\phi$
has to vanish to minimize ${\cal E}_{PE}[\rho_{\pm},\theta, \phi ;R]$.
Then, we find that the configuration minimizing
${\cal E}_{PE}[\rho_{\pm},\theta, \phi=0 ;R]$ with $\rho_{\pm} \neq 0$
is given by
$\rho_{\pm}(y) = \sqrt{\frac{1}{g^2}
(\Lambda^2 - (\frac{\alpha}{R})^2 )}$.
Note here that this configuration is meaningful only when
$\Lambda^2 - (\frac{\alpha}{R})^2 \geq 0$ or
$R \geq \frac{\alpha}{\Lambda}$.
The potential energy for this configuration is given by
\begin{equation}
{\cal E}_{PE}[\rho_{\pm},\theta, \phi=0 ;R]
= \frac{2\pi R \Lambda^4}{g^2}-\frac{2\pi R}{g^2}
\bigg(\Lambda^2 - \bigg(\frac{\alpha}{R}\bigg)^2\bigg)^2
\qquad {\rm for} \quad R \geq \frac{\alpha}{\Lambda}.
\end{equation}
Combining all the results, we conclude that the vacuum configuration
which minimizes ${\cal E}[A_0, A_{\pm};R]$ is given by
\begin{eqnarray}
{\rm for}~~R \leq  R^*\equiv\frac{\alpha}{\Lambda},~
& &\left\{\begin{array}{l}
\vev{A_0(x^{\mu},y)} = {\rm arbitrary~complex~constant},\\
\vev{A_{\pm}(x^{\mu},y)} = 0, \label{u1vac1} \\
\end{array}\right. \\
{\rm for}~~R > R^*\equiv\frac{\alpha}{\Lambda},~
& &\left\{\begin{array}{l}
\vev{A_0(x^{\mu},y)} = 0, \\
\vev{A_{\pm}(x^{\mu},y)} = \frac{1}{g}v\e^{\pm
i\frac{\alpha}{R}(y-y_0)},
\label{u1vac2}\\
\end{array}\right.
\end{eqnarray}
where $v\equiv \sqrt{\Lambda^2-(\alpha/R)^2}$.
The arbitrary parameter $y_0$ reflects
the $U(1)$ symmetry of the theory
and also reflects the fact that the
equations of motion are invariant under the
translation for the $S^1$ direction. The relation between the two
symmetries will be discussed later. Let us note
that when $\alpha = \frac{1}{2}$ we have another choice of the vacuum
configuration $\vev{A_{\pm}(x^{\mu},y)} = \frac{1}{g}
v\e^{\mp i \frac{1}{2R}(y-y_0)}$ for $R > R^*$.
The vacuum in this case is doubly degenerate. One of them corresponds to
the
right moving mode on $S^1$ and the other does to the left moving mode.
\par
As in the $Z_2$ model, there are two phases, depending on the
magnitude of the radius of $S^1$. We see that the vacuum structure
drastically changes at the critical radius $R^*$.
For $R<R^*$, the $U(1)_R$ symmetry is broken, while for $R>R^*$, the
translational invariance for the $S^1$ direction is broken
spontaneously with the breakdown of the $U(1)$ symmetry \cite{mu1}.
The vacuum energy is nonzero for both $R \leq R^*$ and $R > R^*$,
so that
the supersymmetry is broken spontaneously for both regions.
\subsection{Mass spectrum for $R\leq R^*$}
Since we have obtained the vacuum configuration, let us next study
the mass spectrum appeared in $3$-dimensions. The mass spectrum
will be obtained by the same prescription used in the previous section.
Let us start to compute the boson mass spectrum, which can be
read from the quadratic part of the 3-dimensional Lagrangian
\begin{eqnarray}
{\cal L}_{B(R\leq R^*)}^{(2)}=
\int_0^{2\pi R}dy~\Bigl[-\sum_{i=0,\pm}\del_M A_i^*\del^M A_i &+&
\Lambda^2(A_+A_{-}+A_+^* A_{-}^{*}) \nonumber\\
&-&\abs{g\langle A_0 \rangle}^2(\abs{A_+}^2+\abs{A_{-}}^2)
\Bigr],
\label{u1boson}
\end{eqnarray}
where the field $A_0$ has been shifted by the vacuum expectation value
$\langle A_0 \rangle$.
Let us expand the fields $A_0(x, y)$ and
$A_1(x,y)$ in Fourier series for the $S^1$ direction
\begin{equation}
A_0(x,y)=\frac{1}{\sqrt{2\pi R}}\sum_{n\in {\bf Z}}
a_0^{(n)}(x) \e^{i\frac{n}{R}y},\quad
A_{\pm}(x,y)=\frac{1}{\sqrt{2\pi R}}\sum_{n\in {\bf Z}}
a_{\pm}^{(n)}(x) \e^{i\frac{n\pm \alpha}{R}y}.
\end{equation}
These expansions are consistent with the
boundary conditions (\ref{u1bc}).
Inserting these into (\ref{u1boson}) and redefining the
Fourier modes $a_{\pm}^{(n)}$ as
$b_{\pm}^{(n)}\equiv \frac{1}{\sqrt 2}(a_{+}^{(n)}\pm a_{-}^{(-n)*})$,
we have
\begin{equation}
{\cal L}_{B(R\leq R^*)}^{(2)}=
\sum_{n\in {\bf Z}}\Bigl[
-\del_{\mu}a_0^{(n)*}\del^{\mu}a_0^{(n)}-
\sum_{i=\pm}\del_{\mu}b_i^{(n)*}\del^{\mu}b_i^{(n)}
-\Phi_B^{(n)\dagger}({\cal M}_B^{(n)})^2 \Phi_B^{(n)}
\Bigr],
\end{equation}
where $\Phi_B^{(n)}\equiv
(a_0^{(n)},b_+^{(n)},b_{-}^{(n)})^T$
and the squared mass matrix is given by
\begin{equation}
({\cal M}_B^{(n)})^2\equiv
\left(\begin{array}{ccc}
(\frac{n}{R})^2& 0 & 0\\
0 & (\frac{n+\alpha}{R})^2+\abs{g\langle A_0\rangle}^2-\Lambda^2 & 0 \\
0 & 0 & (\frac{n+\alpha}{R})^2+\abs{g\langle A_0\rangle}^2+\Lambda^2 \\
\end{array}\right).
\label{u1bmass}
\end{equation}
Thus, we have obtained the masses of the bosons.
\par
Next, we shall study the fermion mass spectrum for
$R\leq R^* =\alpha/\Lambda$.
It can be read from
\begin{equation}
{\cal L}_{F(R\leq R^*)}^{(2)}=
\int_0^{2\pi R}dy~\Bigl[
-i\sum_{i=0,\pm}{\bar\psi}_i{\bar\sigma}^M\del_M \psi_i
+g\langle A_0 \rangle\psi_+\psi_{-}
+g{\langle A_0 \rangle}^*{\bar\psi}_+{\bar\psi}_{-}\Bigr].
\label{u1fermi}
\end{equation}
We again expand the fields $\psi_0(x,y)$ and $\psi_{\pm}(x,y)$
in Fourier series for the $S^1$ direction
\begin{equation}
\psi_0(x,y)=\frac{1}{\sqrt{2\pi R}}
\sum_{n\in {\bf Z}}\chi_0^{(n)}(x)
\e^{i\frac{n}{R}y},\quad
\psi_{\pm}(x,y)=\frac{1}{\sqrt{2\pi R}}
\sum_{n\in {\bf Z}}\chi_{\pm}^{(n)}(x)
\e^{i\frac{n\pm \alpha}{R}y}.
\label{u1fexp}
\end{equation}
Inserting (\ref{u1fexp}) into (\ref{u1fermi}), we obtain
\begin{equation}
{\cal L}_{F(R\leq R^*)}^{(2)}=\sum_{n \in{\bf Z}}\Bigl[
{\bar\chi}_0^{(n)}(x){\cal M}_{\chi_0}^{(n)}\chi_0^{(n)}(x)
+\half {\bar\Psi}^{(n)}{\cal M}_{\chi_{\pm}}^{(n)}\Psi^{(n)}\Bigr],
\label{u1fermimat}
\end{equation}
where we have defined $\Psi^{(n)}\equiv
({\chi}_+^{(n)},{\bar\chi}_-^{(-n)}{\chi}_-^{(n)},
{\bar \chi}_+^{(-n)})^T$. The matrices ${\cal M}_{\chi_0}^{(n)}$
and ${\cal M}_{\chi_{\pm}}^{(n)}$ in (\ref{u1fermimat}) are
defined by
\begin{eqnarray}
{\cal M}_{\chi_0}^{(n)}&\equiv&
-i{\bar \dslash}+ \frac{n}{R}{\bar \sigma}^y, \\
{\cal M}_{\chi_{\pm}}^{(n)}&\equiv &
\left(\begin{array}{cccc}
-i{\bar\dslash}+\frac{n+\alpha}{R}
{\bar\sigma}^y & g\vev{A_0}^*& 0&0 \\
g\vev{A_0} &-i{\dslash}+\frac{n+\alpha}{R}\sigma^y & 0 &0 \\
0&0&-i{\bar\dslash}+\frac{n-\alpha}{R}{\bar\sigma}^y &
g\vev{A_0}^* \\
0&0&g\vev{A_0}&-i{\dslash}+\frac{n-\alpha}{R}\sigma^y \\
\end{array}\right),
\end{eqnarray}
where $\dslash\equiv \sigma^{\mu}\del_{\mu}$ and $\bar\dslash\equiv
{\bar\sigma}^{\mu}\del_{\mu}$. The mass eigenvalues of the fermions
may be extracted from the determinant of the above matrices, {\it i.e.}
\begin{eqnarray}
\!\!\!\!\!\!{\rm det}{\cal M}_{\chi_0}^{(n)}&=&\del_{\mu}\del^{\mu}
-\biggl(\frac{n}{R}\biggr)^2\ ,\nonumber\\
\!\!\!\!\!\!{\rm det}{\cal M}_{\chi_{\pm}}^{(n)}&=&
\biggl[\del_{\mu}\del^{\mu}-\biggl(\frac{n+\alpha}{R}\biggr)^2
-\abs{g\langle A_0 \rangle}^2\biggr]^2
\biggl[\del_{\mu}\del^{\mu}-\biggl(\frac{n-\alpha}{R}\biggr)^2
-\abs{g\langle A_0 \rangle}^2\biggr]^2.
\end{eqnarray}
Thus, we find
\begin{eqnarray}
(m_{\chi_0^{(n)}})^2 &=& \biggl(\frac{n}{R}\biggr)^2, \nonumber\\
(m_{\chi_{\pm}^{(n)}}^{(n)})^2 &=& \biggl(\frac{n+\alpha}{R}\biggr)^2
+\abs{g\langle A_0 \rangle}^2,
~\biggl(\frac{n-\alpha}{R}\biggr)^2+\abs{g \langle A_0 \rangle}^2.
\label{u1fmass}
\end{eqnarray}
We schematically depict the $R$-dependence
of the boson and fermion mass spectra in the Figure $2$.
\par
\subsection{Analysis of the mass spectrum for $R\leq R^*$}
Let us study here the masses of the bosons and fermions
obtained in the previous subsection.
Note that all the squared masses (\ref{u1bmass})
and (\ref{u1fmass}) are positive
semi-definite for $R\leq \alpha/\Lambda$.
The vacuum configuration (\ref{u1vac1}) is stable for this region.
It is easy to see from the boson mass splitting that the scale of the
SUSY breaking is of order $\Lambda$.
We observe that the mode $a_0^{(0)}$ is massless. The physical
interpretation for this massless mode is that a part of it is guaranteed
by the flatness of the scalar potential at the tree-level and
another part of it corresponds to the Nambu-Goldstone boson
associated with the breakdown of the $U(1)_R$
symmetry due to the vacuum configuration (\ref{u1vac1}).
\par
We also observe that in the mass spectrum there appears a massless
fermion
mode $\chi_0^{(0)}$ which is the Dirac spinor in $3$-dimensions.
This corresponds to the Nambu-Goldstone fermion
associated with the spontaneous SUSY breaking. In order to
confirm this, let us recall the infinitesimal SUSY
transformations for the spinor fields
\begin{eqnarray}
\delta_S\psi_0&=&i\sqrt{2}({\sigma^M{\bar\xi}})\del_M A_0
-\sqrt{2}\xi(\frac{\Lambda^2}{g}-gA_+^*A_{-}^*), \nonumber\\
\delta_S\psi_{\pm}&=&i\sqrt{2}({\sigma^M{\bar\xi}})\del_M A_{\pm}
+\sqrt{2}\xi gA_0^* A_{\mp}^*.
\label{sutra4}
\end{eqnarray}
In the vacuum background given by (\ref{u1vac1}), the vacuum
expectation values of
$\delta_S\psi_0$ and $\delta_S\psi_{\pm}$ become
\begin{equation}
\vev{\delta_S\psi_0}=-\sqrt{2}\xi\frac{\Lambda^2}{g}, \qquad
\vev{\delta_S\psi_{\pm}}=0.
\label{sutra5}
\end{equation}
Thus, for any nontrivial $\xi$, the supersymmetry is completely
broken spontaneously. It follows from (\ref{sutra5}) that
the mode of the Nambu-Goldstone fermion
associated with the SUSY breaking is the one proportional
to $\vev{\delta_S\psi_0}$ in $\psi_0(x,y)$, that is, a constant mode
with respect to $y$ in Fourier series. Therefore, we
confirm that $\chi_0^{(0)}$
is the Nambu-Goldstone fermion associated with the
breakdown of the supersymmetry.
\par
Let us note that the mass spectrum of the bosons and fermions
satisfies the following mass relations for each mode $n$:
\begin{eqnarray}
(m_{a_0^{(n)}})^2+(m_{a_0^{(-n)}})^2
&=& (m_{\chi_0^{(n)}})^2+(m_{\chi_0^{(-n)}})^2, \\
\sum_{i=\pm}\bigg[(m_{b_i^{(n)}})^2+(m_{b_i^{(-n)}})^2\bigg]
&=&
\sum_{i=\pm}\bigg[(m_{\chi_i^{(n)}})^2+(m_{\chi_i^{(-n)}})^2
\bigg].
\label{massmode}
\end{eqnarray}
These are the supertrace formulas for the squared masses in
3-dimensions, though the original proof of the supertrace
formula in Ref.\cite{FGP} does not necessarily imply the
equivalence between the sum of the boson squared masses and that
of the fermion ones for {\it each} Fourier mode.
\subsection{Mass spectrum for $R>R^*$}
In this subsection, we shall study the mass spectrum for the bosons and
fermions for $R > \alpha/\Lambda$. We expand the fields around the
vacuum
configuration (\ref{u1vac2}) and take the quadratic part of
the Lagrangian to obtain the mass spectrum.
The relevant terms for the boson mass spectrum are given by
\begin{eqnarray}
{\cal L}_{B(R> R^*)}^{(2)}&=&
\int_0^{2\pi R}dy~\biggl[
-\sum_{i=0,\pm}\del_M A_i^*\del^M A_i
-v^2\Bigl(2A_0^*A_0+A_{-}^* A_{-}+A_{+}^*A_{+}\nonumber\\
& & +\e^{-i\frac{2\alpha}{R}y} A_{-}^*A_{+}
+\e^{i\frac{2\alpha}{R}y}A_{+}^*A_{-}\Big)
+\biggl(\frac{\alpha}{R}\biggr)^2(A_{+}A_{-}+A_{+}^*A_{-}^*)\biggr],
\label{u1bb}
\end{eqnarray}
where we have put $y_0 =0$ for simplicity.
Let us expand the fields $A_0(x,y)$ and $A_{\pm}(x,y)$ in
Fourier series for the $S^1$ direction as
$$
A_0(x,y)=\frac{1}{\sqrt{2\pi R}}\sum_{n\in {\bf Z}}a_0^{(n)}(x)
\e^{i\frac{n}{R}y},\quad
A_{\pm}(x,y)=\frac{1}{\sqrt{2\pi R}}\sum_{n\in {\bf Z}}a_{\pm}^{(n)}(x)
\e^{i\frac{n\pm \alpha}{R}y},
$$
which are consistent with the boundary conditions (\ref{u1bc}).
Inserting these into (\ref{u1bb}), we obtain
\begin{equation}
{\cal L}_{B(R> R^*)}^{(2)}
=\sum_{n\in {\bf Z}}\half
{\tilde\Phi}_B^{(n)\dagger}(\del_{\mu}\del^{\mu}-({\cal M}_B^{(n)})^2)
{\tilde\Phi}_B^{(n)},
\label{u1bb1}
\end{equation}
where
${\tilde\Phi}_B^{(n)}
\equiv (a_0^{(n)},a_0^{(-n)*},a_{+}^{(n)},a_{-}^{(-n)*},a_{-}^{(n)},
a_{+}^{(-n)*})^T$ and
the squared mass matrix in (\ref{u1bb1}) is given by
\begin{equation}
({\cal M}_{B}^{(n)})^2\equiv \left(\begin{array}{cccccc}
A&0&0&0&0&0 \\
0&A&0&0&0&0\\
0&0&B&E&D&0\\
0&0&E&B&0&D\\
0&0&D&0&C&E\\
0&0&0&D&E&C\\
\end{array}\right),\quad {\rm where}\quad
\begin{array}{l}
A=(\frac{n}{R})^2+2v^2,\\
B=(\frac{n+\alpha}{R})^2+v^2,\\
C=(\frac{n-\alpha}{R})^2+v^2,\\
D=v^2,\\
E=-(\frac{\alpha}{R})^2,
\end{array}
\label{u1bbsq}
\end{equation}
with $v^2\equiv \Lambda^2-(\alpha/R)^2$.
The mass eigenvalues are obtained by solving the equation
${\rm det}(({\cal M}_B^{(n)})^2-m^2{\bf 1}_{6\times 6})=0$.
After the straightforward calculations,
the eigenvalues are found to be
\begin{eqnarray}
m^2
&=&\left\{\begin{array}{l}
2v^2+(\frac{n}{R})^2,\\
2v^2+(\frac{n}{R})^2,\\
v^2+(\frac{n}{R})^2
\pm\sqrt{v^4
+4(\frac{\alpha}{R})^2(\frac{n}{R})^2}, \\
v^2+2(\frac{\alpha}{R})^2+(\frac{n}{R})^2
\pm\sqrt{v^4
+4(\frac{\alpha}{R})^2(\frac{n}{R})^2}. \\
\end{array}\right.
\label{u1beigen}
\end{eqnarray}
\par
Let us proceed to compute the fermion mass spectrum.
It can be read from
\begin{eqnarray}
{\cal L}_{F(R> R^*)}^{(2)}=
\int_0^{2\pi R}dy~\Bigl[
-i\sum_{i=0,\pm}{\bar\psi}_i{\bar\sigma}^M\del_M\psi_i
&+&v\psi_0(\e^{-i\frac{\alpha}{R}y}
\psi_{+}+\e^{i\frac{\alpha}{R}y}\psi_{-}) \nonumber \\
&+&v{\bar\psi}_0(\e^{i\frac{\alpha}{R}y}
{\bar\psi}_{+}+\e^{-i\frac{\alpha}{R}y}{\bar\psi}_{-})
\Bigr].
\end{eqnarray}
We find that it is convenient to redefine the spinor fields so as to
satisfy the periodic boundary condition, {\it i.e.}
$\psi_{\pm}(x,y)\equiv
\e^{\pm\frac{i\alpha}{R}y}{\tilde\psi}_{\pm}(x,y)$,
where ${\tilde\psi}_{\pm}(x,y+2\pi R)={\tilde\psi}_{\pm}(x,y)$.
Moreover, we make use of the prescription (\ref{diracs}) given
in the Appendix B. Thus, we obtain
\begin{equation}
{\cal L}_{F(R> R^*)}^{(2)}=
\int_0^{2\pi R}dy~{\bar\Psi}(-i\gamma^{\mu}\del_{\mu}+{\cal M}_F)\Psi,
\label{u1ff}
\end{equation}
where $\Psi\equiv (\psi^c_0,{\tilde\psi}_{+},{\tilde\psi}_{-})^T$
and $\psi^c_0$ denotes the charge conjugation of $\psi_0$.
Note that the spinors in above equation are the Dirac spinors in
$3$-dimensions.
The mass matrix ${\cal M}_F$ in (\ref{u1ff}) is given by
\begin{equation}
{\cal M}_F=\left(\begin{array}{ccc}
-i\del_y& -iv& -iv\\
iv&i\del_y-\alpha/R& 0 \\
iv&0&i\del_y+\alpha/R\\
\end{array}\right).
\label{u1ffm}
\end{equation}
We expand the spinor fields in Fourier series for the $S^1$ direction
by taking account of the boundary
condition for $\psi^c_0$ and ${\tilde\psi}_{\pm}$ as
\begin{equation}
\psi_0^c(x,y)=\frac{1}{\sqrt{2\pi R}}\sum_{n\in {\bf Z}}\chi_0^{(n)}(x)
\e^{i\frac{n}{R}y}, \quad
{\tilde\psi}_{\pm}(x,y)
=\frac{1}{\sqrt{2\pi R}}\sum_{n\in {\bf Z}}\chi_{\pm}^{(n)}(x)
\e^{i\frac{n}{R}y}.
\end{equation}
Then, we find from (\ref{u1ff})
\begin{equation}
{\cal L}_{F(R> R^*)}^{(2)}
=\sum_{n\in {\bf Z}}
({\bar\chi}_0^{(n)},{\bar\chi}_{+}^{(n)},{\bar\chi}_{-}^{(n)})
(-i\gamma^{\mu}\del_{\mu}+{\cal M}_F^{(n)})
\left(\begin{array}{c}
\chi_0^{(n)}\\
\chi_{+}^{(n)}\\
\chi_{-}^{(n)}
\end{array}\right),
\end{equation}
where
\begin{equation}
{\cal M}_F^{(n)}=({\cal M}_F^{(n)})^{\dagger}=\left(\begin{array}{ccc}
n/R&-iv&-iv\\
iv&-(n+\alpha)/R&0\\
iv& 0& -(n-\alpha)/R\\
\end{array}\right).
\end{equation}
To obtain the mass spectrum, it is sufficient to
diagonalize the square of ${\cal M}_F^{(n)}$.
We finally have
\begin{equation}
({\cal M}_F^{(n)})^2=\left(\begin{array}{ccc}
(n/R)^2+2v^2&iv\frac{\alpha}{R}&-iv\frac{\alpha}{R}\\
-iv\frac{\alpha}{R}&((n+\alpha)/R)^2+v^2&v^2\\
iv\frac{\alpha}{R}& v^2& ((n-\alpha)/R)^2+v^2\\
\end{array}\right).
\label{u1ffsq}
\end{equation}
We shall not here try to solve the eigenvalue equation for the
fermion masses exactly, though the mass eigenvalues could be
obtained by solving a cubic equation. However, some of important
properties, which will be discussed in the next subsection,
can easily be extracted from the expression (\ref{u1ffsq}).
The schematic behavior of the boson and fermion masses
with respect to $R$ is found in the Figure $2$.
\subsection{Analysis of the mass spectrum for $R>R^*$}
It is easy to see from (\ref{u1beigen}) that the squared
masses for the bosons are all positive semi-definite
for $R>\alpha/\Lambda$ with $0<\alpha \leq \half$.
Since the vacuum configuration (\ref{u1vac2})
breaks both the translational
invariance for the $S^1$ direction and
the global $U(1)$ symmetry, one might expect {\it two}
massless Nambu-Goldstone modes associated with the broken
generators of the symmetries.
If we take a look at the boson mass spectrum, however, there is only
one massless mode given by
\begin{equation}
m^2=v^2+(\frac{n}{R})^2
- \sqrt{v^4
+4(\frac{\alpha}{R})^2(\frac{n}{R})^2}=0\qquad{\rm for}\quad n=0.
\end{equation}
A physical interpretation for this goes as follows: Let us
consider the $U(1)'$ transformation, which is a linear
combination of the translation for the $S^1$ direction and
the $U(1)$ transformation, defined by
\begin{equation}
U(1)': \quad A_0(x,y) \rightarrow A_0(x, y+a), \quad
A_{\pm}(x,y) \rightarrow
e^{\mp i\frac{\alpha}{R}a}A_{\pm}(x, y+a).
\label{u1dash}
\end{equation}
It turns out that the $U(1)'$ transformation is an exact symmetry
of the model. In fact, the vacuum configuration (\ref{u1vac2}) is
invariant under the modified $U(1)'$ transformation (\ref{u1dash}).
Therefore, one linear combination of the two symmetry generators
survives as an unbroken generator and the massless boson
corresponds to the Nambu-Goldstone mode associated with the
broken generator.
\par
It is easy to see that there is a massless mode
(in the sense of the 3-dimensional Dirac spinor)
in the fermion mass
spectrum, which can be seen from the fact that the
determinant of the matrix (\ref{u1ffsq}) can be zero.
The determinant of the mass matrix is computed as
${\rm det}({\cal M}_F^{(n)})^2=(\frac{n}{R})^2((\frac{n}{R})^2
-(\frac{\alpha}{R})^2+2v^2)^2$.
Since $v^2>0$ and $n^2-\alpha^2>0$ for $n\neq 0$, the determinant
of the matrix can be zero only when $n=0$. This implies that
there is a massless eigenvalue
in the mass matrix. This massless mode corresponds to the
Nambu-Goldstone fermion associated with the
spontaneous SUSY breaking. To see this, let us consider the
vacuum expectation values of
the SUSY transformations for the spinors
in the vacuum configuration (\ref{u1vac2}), {\it i.e.}
\begin{equation}
\vev{\delta_S\psi_0}=-\frac{\sqrt{2}}{g}\xi
\bigg(\frac{\alpha}{R}\bigg)^2,
\quad
\vev{\delta_S\psi_{\pm}}=\frac{\sqrt{2}}{g}i(\sigma^y{\bar\xi})
v\bigg(\pm i\frac{\alpha}{R}\bigg)
\e^{\pm i\frac{\alpha}{R}y}.
\label{susytra4}
\end{equation}
Since $\vev{\delta_S\psi_0}$ and $\vev{\delta_S\psi_{\pm}}$ do not
vanish for any nontrivial choice of $\xi$ and $\bar \xi$, the
supersymmetry is spontaneously broken completely. The corresponding
Nambu-Goldstone (3-dimensional Dirac) fermion is the massless mode
found above.
\par
In the $Z_2$ model, we have not found any simple relations between
the boson and the fermion masses for $R > R^*$. In the $U(1)$ model,
we have, however, found that the mass matrices (\ref{u1bbsq}) and
(\ref{u1ffsq}) lead to the supertrace formula which holds for each
Fourier mode between the bosons and the fermions, {\it i.e.}
${\rm Tr}({\cal M}_B^{(n)})^2 = {\rm Tr}({\cal M}_F^{(n)})^2$.
\par
It is interesting to study the supersymmetry breaking in the
$R \rightarrow \infty$ limit. In this limit, the supersymmetry
transformations for the spinor fields vanish as seen from
(\ref{susytra4}). Therefore, the supersymmetry is restored
in this limit, which is very contrary to the $Z_2$ model in which
the supersymmetry is broken even in this limit. The net effect
of the boundary conditions of the $U(1)$ model completely
disappears and we will have the $N=2$ supersymmetry in 3-dimensions
in the limit of $R \rightarrow \infty$. Finally, let us summarize
the phase structure of the $U(1)$ model in the Table IV-1.
\section{CONCLUSIONS AND DISCUSSIONS}
In this paper we have proposed a new mechanism of the spontaneous
SUSY breaking. We have discussed the general criteria
for the superpotential in order for our mechanism to work.
We have obtained the two minimal models, $Z_2$ and $U(1)$ models
to realize our mechanism and studied the dynamical aspects,
such as the vacuum structure and the mass spectra in the models.
\par
The crucial point of our
mechanism is that there exist
the solutions to the $F$-term conditions but they are not realized as
the vacuum configurations, as contrary to the usual cases, because of
the nontrivial boundary conditions of the fields for the
compactified direction. It is important to investigate the ``energy''
functional including the kinetic terms of the scalar fields
in order to find the true vacuum configuration consistent
with the nontrivial boundary conditions.
A remarkable observation in our analysis is that the vacuum
configurations which depends on the coordinate of the extra
dimensions are energetically favorable than
any coordinate-independent configurations for $R > R^*$.
In other words, there exists
a phase in which the translational
invariance of the compactified direction is spontaneously broken when
the size of the extra dimension exceeds the critical radius, at which
the phase transition occurs.
On the other hand, the supersymmetry is broken spontaneously
irrespective of the size of the extra dimension in both models.
\par
The complexity of the vacuum structures
reflects the mass spectra of
the models, which have rich structures and depend
on the size of the extra dimension in a nontrivial way.
The vacuum configuration breaks some of the symmetries of the
models, so that the Nambu-Goldstone bosons and/or fermions associated
with the breakdown of the symmetries appear. We have actually
observed those massless modes in the mass spectra.
\par
Our mechanism can cause quite different vacuum structures if the
models have different global symmetries
whose degrees of freedom are available
to impose nontrivial boundary conditions on the fields.
This is easily observed in
the two models we have studied.
The vacuum structures of the two models are quite different,
as seen from (\ref{zvac1}), (\ref{zvac2}), (\ref{u1vac1})
and (\ref{u1vac2}). Moreover, the behavior of the models in the
limit of $R\rightarrow \infty$ is also quite different.
In the $Z_2$ model the vacuum configuration is reduced to a single kink
solution, which is one of the topologically stable solutions
\begin{equation}
{\vev{A_1(x^{\mu},y)}|}_{R=\infty}=\frac{\sqrt{2}\Lambda}{g}
\tanh\bigg(\frac{\Lambda}{\sqrt 2}y\bigg).
\end{equation}
The supersymmetry is still broken in this limit because
the vacuum expectation values of the SUSY transformations for
$\psi_0$ and $\psi_1$ do not vanish even in this background
\begin{equation}
\vev{\delta_S \psi_0}|_{R\rightarrow \infty}
=-\sqrt{2}\frac{\Lambda^2}{g}\xi\frac{1}{\cosh^2(\frac{\Lambda y}{\sqrt
2})},
\quad
\vev{\delta_S \psi_1}|_{R\rightarrow \infty}
=i\sqrt{2}\frac{\Lambda^2}{g}\sigma^y{\bar\xi}
\frac{1}{\cosh^2(\frac{\Lambda y}{\sqrt 2})}.
\\ \nonumber
\label{sutra3}
\end{equation}
For any nontrivial $\xi$, there is no SUSY transformation
which acts trivially on the kink solution. Therefore,
the supersymmetry is
still broken in the limit of
$R \rightarrow \infty$ to have no SUSY in
$3$-dimensions \cite{shif}, while in the $U(1)$ model,
the supersymmetry is
restored in the limit of $R \rightarrow \infty$
to have the $N=2$ SUSY in $3$-dimensions.
This is easily seen from (\ref{susytra4}).
This is very contrary to the $Z_2$ model in which
the supersymmetry is
still completely broken in this limit.
The effect of the boundary conditions on the fields
completely disappears to be able to have a constant
vacuum expectation value
of $\vev{A_1}$. These are direct consequences followed from the
fact that in the two models the vacuum structures
have quite different dependence
on the coordinate of the extra dimension.
These results will not be specific to the
models considered in this paper,
but are expected to be general ones of
the theories in which our mechanism works.
\par
Let us comment on an effective theory appeared in $3$-dimensions.
If one tries to construct the effective theory in $3$-dimensions by
only massless modes, we have free field theories for the $Z_2$ and
$U(1)$ models in the region of $R\leq R^*$. On the other hand, in the
region of $R> R^*$, we have interacting field theories with
Yukawa couplings in the two models. In the limit of $R\rightarrow
\infty$,
the effective $Z_2$ model has no
supersymmetry, while the $U(1)$ model has the 3-dimensional
$N=2$ supersymmetry.
\par
There will be many ways to extend our work.
It may be interesting to ask how our mechanism works on more complex
manifolds, such as torus. We expect that there will appear complicated
phase structures, depending on the size of their compactified spaces
and on how we impose nontrivial boundary conditions on superfields.
We can also study models with nonabelian global symmetries
such as $SU(N)$, instead of $Z_2$ and $U(1)$ symmetries.
In addition to them, it is important to study gauge theories
and to see how our mechanism works and
what new dynamics are hidden in them.
It may also be interesting to investigate how the partial SUSY
breaking occurs in the gauge theory in the connection of the
well-known BPS objects. These will be reported in the near future.
\newpage
\begin{center}
{\bf Acknowledgement}
\end{center}
\vspace{12pt}
M. T. was supported by Grant-in-Aid Scientific Research, Grant No.3666.
K. T. would like to thank The Niels Bohr Institute and
INFN, Sezione di Pisa for warm hospitality. 
\vskip 1cm
\noindent{\large{\bf APPENDIX A}}
\vskip 0.5cm
In this appendix, we show how we determine the vacuum configuration
in the $Z_2$ model in detail.
The vacuum configuration $\vev{A_0}$ and $\vev{A_1}$ have to be stable
against any infinitesimal variations of $A_0$ and $A_1$. So, they have
to
satisfy the field equations
derived from the ``energy'' functional (\ref{efunc})
\begin{eqnarray}
0=\frac{\delta{\cal E}[\vev{A_0},\vev{A_1}]}{\delta A_0^*(y)}
&=&-\frac{d^2\vev{A_0(y)}}{dy^2}+g(g\vev{A_0(y)}-\mu)\abs{\vev{A_1(y)}}^2,
\label{eom1}\\
0=\frac{\delta{\cal E}[\vev{A_0},\vev{A_1}]}{\delta A_1^*(y)}
&=&-\frac{d^2\vev{A_1(y)}}{dy^2}-g\vev{A_1^*(y)}
\bigg(\frac{\Lambda^2}{g}-
\frac{g}{2}\vev{A_1(y)}^2\bigg) \nonumber\\
&+&\abs{g\vev{A_0(y)}-\mu}^2 \vev{A_1(y)}^2.
\label{eom2}
\end{eqnarray}
It is convenient to separately investigate the two cases:
Type~(I);$\langle A_1(y)\rangle = 0$ and
type~(II);$\langle A_1(y)\rangle \neq 0$.
Let us first consider the type (I) solution. In this case we find
immediately
\begin{equation}
{\cal E}[A_0, A_1=0] = \int_{0}^{2\pi R}dy\left[
\bigg|\frac{dA_0}{dy}\bigg|^2 + \frac{\Lambda^4}{g^2} \right ] \geq
\frac{2\pi R \Lambda^4}{g^2}.
\end{equation}
The equality holds only when
$\frac{dA_0}{dy} = 0$ or equivalently
$A_0(y) ={\rm arbitrary~complex~constant}$.
Thus, the type (I) solution is found to be
$A_0^{({\rm I})}(y)={\rm arbitrary~complex~constant}$ and
$A_1^{({\rm I})}(y)=0$.
It is obvious that these satisfy the
equations of motion (\ref{eom1}) and (\ref{eom2}).
For the type (I) solution the ``energy'' functional is given by
${\cal E}[A_0^{({\rm I})}, A_1^{({\rm I})}]
= \frac{2\pi R \Lambda^4}{g^2}$.
\par
We shall next consider the type (II) solution. Then, we may rewrite
the ``energy'' functional ${\cal E}[A_0, A_1]$ as
\begin{equation}
{\cal E}[A_0^{({\rm II})}, A_1^{({\rm II})}]
= {\cal E}[A_1^{({\rm II})}]
+ \Delta{\cal E}[A_0^{({\rm II})}, A_1^{({\rm II})}],
\label{newefunc}
\end{equation}
where
\begin{eqnarray}
{\cal E}[A_1^{({\rm II})}] &\equiv& \int_0^{2\pi R}dy \left[
\bigg|\frac{dA_1^{({\rm II})}}{dy}\bigg|^2
+ \bigg|\frac{\Lambda^2}{g}
- \frac{g}{2}(A_1^{({\rm II})})^2\bigg|^2 \right],
\\
\Delta{\cal E}[A_0^{({\rm II})}, A_1^{({\rm II})}]
&\equiv& \int_0^{2\pi R}dy
\left[\bigg|\frac{dA_0^{({\rm II})}}{dy}\bigg|^2
+ |gA_0^{({\rm II})}-\mu|^2|A_1^{({\rm II})}|^2 \right].
\end{eqnarray}
Obviously, $\Delta{\cal E}[A_0^{({\rm II})}, A_1^{({\rm II})}]$
is positive semi-definite.
The equality $\Delta{\cal E}[A_0^{({\rm II})}, A_1^{({\rm II})}]=0$
holds only when $A_0^{({\rm II})}=\mu/g$.
Since we are interested in the vacuum configuration, it turns out that
it is sufficient to consider the
minimization problem of ${\cal E}[A_1^{{\rm (II)}}]$, instead of
${\cal E}[A_0^{{\rm (II)}}, A_1^{{\rm (II)}}]$,
with $A_0^{({\rm II})}=\mu/g$.
\par
It is useful to parametrize the field
$A_1^{{\rm (II)}}(y)$ as
$A_1^{{\rm (II)}}(y) \equiv
\frac{1}{\sqrt{2}}\left(\xi(y)+i\eta(y)\right)$,
where $\xi(y)$ and $\eta(y)$ are real fields satisfying
$\xi(y+2\pi R) = -\xi(y)$ and
$\eta(y+2\pi R) = -\eta(y)$.
Inserting these representations into
${\cal E}[A_1^{({\rm II})}]$, we find
\begin{equation}
{\cal E}[A_1^{{\rm (II)}}] = {\cal E}[\xi]
+ \Delta{\cal E}'[\xi, \eta],
\end{equation}
where
\begin{eqnarray}
{\cal E}[\xi] &\equiv& \int_0^{2\pi R}dy\left[\frac{1}{2}
\bigg(\frac{d\xi}{dy}\bigg)^2
+ \bigg(\frac{\Lambda^2}{g}-\frac{g}{4}\xi^2\bigg)^2\right ],
\\
\Delta{\cal E}'[\xi, \eta] &\equiv& \int_0^{2\pi R}dy \left[
\frac{1}{2}\bigg(\frac{d\eta}{dy}\bigg)^2
+\frac{\Lambda^2}{2}\eta^2 +
\frac{g^2}{8}\xi^2 \eta^2 +\frac{g^2}{16}\eta^4 \right ].
\end{eqnarray}
Obviously, $\Delta{\cal E}'[\xi, \eta]$ is positive semi-definite
and the equality $\Delta{\cal E}'[\xi, \eta]=0$ holds only when
$\eta(y)=0$.
Therefore, we have found that the type (II) solution for the
vacuum configuration should be of the form,
$A_0^{{\rm (II)}} = \mu/g$ and
$A_1^{{\rm (II)}} = \frac{1}{\sqrt{2}}\xi(y)$.
Now the minimization problem of
${\cal E}[A_0^{{\rm (II)}}, A_1^{{\rm (II)}}]$
is equivalent to that of ${\cal E}[\xi]$
with the boundary condition
\begin{equation}
\xi(y+2\pi R) = - \xi(y).
\label{bcrf}
\end{equation}
Here, any configurations of $\xi(y)$ which minimize ${\cal E}[\xi]$
have to satisfy the equation of motion
\begin{equation}
0 = \frac{\delta {\cal E}[\xi]}{\delta \xi(y)}=
- \frac{d^2 \xi(y)}{dy^2}
-g\xi\bigg(\frac{\Lambda^2}{g}-\frac{g}{4}\xi^2\bigg).
\label{eomrf}
\end{equation}
It is important to note that if there exists any nontrivial type (II)
solution, then, the type (I) solution is no longer the vacuum
configuration.
Actually, using (\ref{eomrf}), we can easily show that
\begin{equation}
{\cal E}[A_0^{{\rm (II)}}, A_1^{{\rm (II)}}] =
\frac{2\pi R \Lambda^4}{g^2} - \int_0^{2\pi R} dy
\frac{g^2}{16}(\xi(y))^4 \leq {\cal E}[A_0^{({\rm I})}, A_1^{({\rm
I})}].
\label{econst}
\end{equation}
General solutions satisfying (\ref{eomrf}) are known to be given by
\begin{equation}
\xi(y) = \frac{2\sqrt{2}k \omega}{g} {\rm sn}(\omega(y-y_0), k)
\quad {\rm with}\quad \omega\equiv \frac{\Lambda}{\sqrt{1+k^2}}.
\end{equation}
Here, sn$(u,k)$ is the Jacobi elliptic function
whose period is given by
$4K(k)$, where $K(k)$ denotes the complete
elliptic function of the first kind defined by
\begin{equation}
K(k)\equiv \int_0^1\frac{dx}{\sqrt{(1-x^2)(1-k^2x^2)}}.
\end{equation}
The parameters  $k~ (0\leq k<1)$ and $y_0$ are integration constants.
Since we have found the general solutions to the equation
of motion (\ref{eomrf}) (precisely speaking, we find the general
solutions for periodic motions), the next task is to extract desired
solutions from the general solutions, which have to satisfy the
boundary condition (\ref{bcrf}). It leads to
${\rm sn}(\omega(y+2\pi R-y_0),k)=-{\rm sn}(\omega(y-y_0),k)$.
This relation can be satisfied provided that
\begin{equation}
2\pi R\omega = (2n-1)2K(k)
\label{constraint}
\end{equation}
for some positive integer $n$. It should be emphasized that there
do {\it not} always exist solutions
for any values of $R$ and $n$. Indeed,
in order for a solution of (\ref{constraint}) to exist, the
inequality $R\geq \frac{n-\half}{\Lambda}$ has to be satisfied,
where we have used $K(k)\geq K(0)=\pi/2$ and $\sqrt{1+k^2}\geq 1$.
\par
We are, now, in a position to decide which configuration
minimizes the ``energy'' functional (\ref{newefunc}). In the
above analysis, we have
found that the vacuum configuration should be given by one of
the configurations
\begin{eqnarray}
{\rm type (I)};~~& &\left\{\begin{array}{l}
A_0^{({\rm I})}(y)={\rm arbitrary~complex~constant}, \nonumber \\
A_1^{({\rm I})}(y)=0. \end{array}\right. \\
{\rm type (II)};~~& &
\left\{\begin{array}{l}
A_0^{({\rm II},n)}(y)=\mu/g, \nonumber \\
A_1^{({\rm II},n)}(y)=\frac{2k\omega}{g}~{\rm sn}(\omega(y-y_0),k),
\end{array}\right.
\end{eqnarray}
where $n\in {\bf Z}>0$ and $k$ is determined by
the condition (\ref{constraint}).
Let us decide which configuration is the vacuum one.
\par
In the region of $0<R\leq 1/{2\Lambda}$, the
condition (\ref{constraint}) cannot be satisfied
for any $n$. So, there exists
only the type (I) solution. We immediately conclude that the type (I)
solution is the vacuum configuration in this region. The ``energy''
functional is then given by
${\cal E}[A_0^{({\rm I})}, A_1^{({\rm I})}]
=\frac{2\pi R\Lambda^4}{g^2}$.
In the region of $1/{2\Lambda}< R\leq 3/{2\Lambda}$, we have the
type (II) solution with $n=1$ as well as the
type (I) solution. As we have already shown in (\ref{econst}),
the vacuum configuration
should be given by the type (II) solution with $n=1$.
In the same way, for the region of $({N-1/2})/{\Lambda}<
R\leq ({N+1/2})/{\Lambda}$,  we have the
type (II) solutions with $n=1,2,\cdots, N$ as well as the
type (I) solution. The ``energy'' functional for each solution is given
by
\begin{eqnarray}
{\cal E}[A_0^{({\rm I})},A_1^{({\rm I})}]&=&
\frac{2\pi R \Lambda^4}{g^2}, \\
{\cal E}[A_0^{({\rm II},n)},A_1^{({\rm II},n)}]&=&
\frac{2(2n-1)\Lambda^{3}}{3(1+k^2)^{\frac{3}{2}}g^{2}}
\left[-(1-k^2)(5+3k^2)K(k)+8(1+k^2)E(k)\right],
\end{eqnarray}
where $E(k)$ is the complete elliptic function of the second kind
and $k$ is determined by the relation (\ref{constraint}).
Then, it is straightforward to show that
\begin{equation}
\frac{d{\cal E}[A_0^{({\rm II},n)}, A_1^{({\rm II},n)}]}{dR}
= \frac{2\pi \Lambda^4}{g^2}
\left( \frac{1-k^2}{1+k^2} \right )^2 \geq 0.
\end{equation}
It follows that ${\cal E}[A_0^{({\rm II},n)}, A_1^{({\rm II},n)}]$ is a
monotonically increasing function of $R$. Furthermore, we can
show that
${\cal E}[A_0^{({\rm II},n)},
A_1^{({\rm II},n)}]|_{R=\frac{(n-\half)}{\Lambda}}
\leq {\cal E}[A_0^{({\rm II},n)}, A_1^{({\rm II},n)}]|_R
\leq {\cal E}[A_0^{({\rm II},n)}, A_1^{({\rm II},n)}]|_{R=\infty}$,
which means
$(2n-1)\frac{\pi \Lambda^3}{g^2}
\leq {\cal E}[A_0^{({\rm II},n)}, A_1^{({\rm II},n)}]|_R
\leq (2n-1)\frac{8\sqrt{2}\Lambda^3}{3g^2}$.
The $R$-dependence of ${\cal E}[A_0,A_1]$ is depicted
in the Figure $3$. Therefore, we can easily show that
${\cal E}[A_0^{({\rm II},1)}, A_1^{({\rm II},1)}]
\leq {\cal E}[A_0^{({\rm II},n)}, A_1^{({\rm II},n)}]
< {\cal E}[A_0^{({\rm I})}, A_1^{({\rm I})}]$.
This inequalities mean that for $R >1/{2\Lambda}$ the vacuum
configuration is given by
$\vev{A_0(y)}=A_0^{({\rm II},1)}(y)$ and
$\vev{A_1(y)}=A_1^{({\rm II},1)}(y)$.
Hence, we obtain the vacuum configuration of the $Z_2$-model
as (\ref{zvac1}) and (\ref{zvac2}) given in the text.
\par
The condition (\ref{constraint}) gives
\begin{equation}
2\pi R\omega=2K(k)
\label{newconst}
\end{equation}
for the true vacuum configuration. It is also easy to see  from
the condition (\ref{newconst}) that
the critical radius $R^*$ is given by
$R=\frac{1}{\pi\omega}K(k)\geq \frac{K(0)}{\pi
\Lambda}=\frac{1}{2\Lambda}
\equiv R^*$,
which is nothing but the critical radius obtained in the text.
\vskip 0.5cm
\noindent{\large{\bf APPENDIX B}}
\vskip 0.3cm
\noindent{\bf B.1 $2$-component spinors in $3$-dimensions
from $4$-dimensions}
\vskip 0.2cm

Let us summarize the relations between 4-dimensional gamma matrices
and 3-dimensional ones.
We also present the decompositions
of 4-dimensional spinors into 3-dimensional ones.
\par
Let us define the $\sigma$-matrices in $4$-dimensions by
\begin{equation}
\begin{array}{ll}
\sigma^0=-{\bf 1}_{2\times 2}={\bar\sigma}^0,&
\sigma^1=\tau_z=-{\bar\sigma}^1,\\
\sigma^2=\tau_x=-{\bar\sigma}^2,&
\sigma^{3}=\tau_y=-{\bar\sigma}^{3},
\quad ({\rm or} \quad
\sigma^{y}=\tau_y=-{\bar\sigma}^{y}),
\end{array}
\end{equation}
where $\tau_i (i=x,y,z)$ is the Pauli matrix.
Then, we define the $\gamma$-matrices in $3$-dimensions by
$\gamma^{\mu}\equiv \tau_y{\bar\sigma}^{\mu}$$(\mu=0,1,2)$.
The $\gamma$-matrices in $3$-dimensions satisfy
$\{\gamma^{\mu},\gamma^{\nu}\}=-2\eta^{\mu\nu}$
with diag~$(\eta^{\mu\nu})=(-,+,+)$,
and they have properties such as
\begin{equation}
\begin{array}{ccc}
(\gamma^0)^{\dagger}=\gamma^0,& (\gamma^0)^T=-\gamma^0,
&(\gamma^0)^*=-\gamma^0,\\ [2mm]
(\gamma^1)^{\dagger}=-\gamma^1,& (\gamma^1)^T=\gamma^1,
&(\gamma^1)^*=-\gamma^1,\\ [2mm]
(\gamma^2)^{\dagger}=-\gamma^2,& (\gamma^2)^T=\gamma^2,
&(\gamma^2)^*=-\gamma^2.
\end{array}
\end{equation}
This representation is the Majorana representation.
The charge conjugation in $3$-dimensions is defined by
$\psi^c\equiv C{\bar\psi}^T$,
where ${\bar\psi}=\psi^{\dagger}\gamma_0=-\psi^{\dagger}\gamma^0$.
The $C$ is the charge conjugation matrix defined by
$C\equiv\gamma^0=-\gamma_0=-\tau_y$.
A Majorana fermion in $3$-dimensions is defined by
$\psi=\psi^c=\psi^*$.
According to the definitions we made above, we have
the prescription used in the text.
Let us denote a $2$-component Dirac spinor in $3$-dimensions and
a $2$-component Weyl spinor in $4$-dimensions by
$\psi$ and $\psi_{(4)}$, respectively. Then, we have
\begin{equation}
\psi_{(4)}^{\alpha}
=(i\tau_y\psi)_{\alpha}=(-i\psi^T\tau_y)_{\alpha},\quad
{\bar\psi}_{(4)}^{\dot\alpha}
=(i\tau_y\psi^*)_{\alpha}=(-i\psi^{\dagger}\tau_y)_{\alpha}.
\end{equation}
\begin{equation}
\begin{array}{cc}
\psi_{(4)}^{\alpha}\chi_{(4)\alpha}
=-i{\bar{\psi^c}}\chi,&
{\bar\chi}_{(4)\dot\alpha}{\bar\psi}_{(4)}^{\dot\alpha}
=i{\bar\chi}\psi^c,\\
{\bar\psi}_{(4)\dot\alpha}({\bar\sigma}^{\mu})^{\dot\alpha\beta}
\chi_{(4)\beta}={\bar\psi}\gamma^{\mu}\chi, &
\chi_{(4)}^{\alpha}({\sigma}^{\mu})_{\alpha \dot\beta}
{\bar\psi}_{(4)}^{\dot\beta}={\bar\chi}^c\gamma^{\mu}\psi^c, \\
{\bar\psi}_{(4)\dot\alpha}({\bar\sigma}^{y})^{\dot\alpha\beta}
\chi_{(4)\beta}=-{\bar\psi}\chi, &
\chi_{(4)}^{\alpha}({\sigma}^{y})_{\alpha \dot\beta}
{\bar\psi}_{(4)}^{\dot\beta}={\bar\chi}^c\psi^c.
\label{diracs}
\end{array}
\end{equation}
It is useful that
we express further a Dirac spinor in terms of
two Majorana spinors in $3$-dimensions. Let $\psi$ and $\chi$
be two 3-dimensional Dirac spinors. They can be expressed in
terms of 3-dimensional Majorana spinors such as
\begin{equation}
\psi\equiv \frac{1}{\sqrt 2}(\rho_1+i\rho_2),\qquad
\chi\equiv \frac{1}{\sqrt 2}(\xi_1+i\xi_2),
\label{majo}
\end{equation}
where $\rho_1$, $\rho_2$, $\xi_1$ and $\xi_2$ are the Majorana
spinors in 3-dimensions.
The 3-dimensional Majorana spinors used in the text are defined
through (\ref{majo}).
\vskip 0.5cm
\noindent{\bf B.2 Eigenvalues and eigenfunctions of the
Lam\'e equation}
\vskip 0.2cm

In this subsection, we briefly summarize
general properties of the Lam\'e equation
\begin{equation}
\Bigl[-\frac{d^2}{du^2}+N(N+1)k^2~{\rm sn}^2(u,k)\Bigr]\phi^{(i)}(u,k)
=\Omega^{(i)}(k)\phi^{(i)}(u,k).
\end{equation}
The solutions to
the Lam\'e equation with
$\phi(u+2K(k))=\pm \phi(u)$ are known to be classified by the
four types of the eigenfunctions denoted by
$Ec_N^{2n}(u,k), Ec_N^{2n+1}(u,k), Es_N^{2n+2}(u,k)$ and
$Es_N^{2n+1}(u,k)$
with $n=0, 1, 2, \cdots$ \cite{Lame}. These satisfy
\begin{eqnarray}
Ec_N^{2n}(-u,k)&=&Ec_N^{2n}(u,k), \quad
Ec_N^{2n}(u+2K(k),k)=Ec_N^{2n}(u,k), \nonumber \\ [2mm]
Ec_N^{2n+1}(-u,k)&=&Ec_n^{2n+1}(u,k), \quad
Ec_N^{2n+1}(u+2K(k),k)=-Ec_N^{2n+1}(u,k), \nonumber \\ [2mm]
Es_N^{2n+2}(-u,k)&=&-Es_N^{2n+2}(u,k), \quad
Es_N^{2n+2}(u+2K(k),k) = Es_N^{2n+2}(u,k), \nonumber  \\ [2mm]
Es_N^{2n+1}(-u,k)&=&-Es_N^{2n+1}(u,k), \quad
E_N^{2n+1}(u+2K(k),k) = -Es_N^{2n+1}(u,k).
\end{eqnarray}
Let us denote the eigenvalues belonging to $Ec_N^n(u,k)$ and
$Es_N^n(u,k)$ by
$\alpha_N^n(k)$ and $\beta_N^n(k)$, respectively.
Since the integer $n$ corresponds to the number of nodes of the
eigenfunctions, we have the following increasing sequence
of the eigenvalues:
\begin{eqnarray}
\alpha_N^0 < &\alpha_N^1& < \alpha_N^2 < \cdots,\nonumber \\
\beta_N^1 < &\beta_N^2& < \beta_N^3 < \cdots,\nonumber\\
\alpha_N^1 < &\beta_N^2& < \alpha_N^3 < \beta_N^4 < \cdots,\nonumber \\
\alpha_N^0 < &\beta_N^1& < \alpha_N^2 < \beta_N^3 < \cdots.
\label{hieboson}
\end{eqnarray}
The following results about the degeneracy of the eigenvalue
have been known:
\begin{itemize}
\item $\alpha_N^n \neq \beta_N^n$ for all $n=0, 1, 2, \cdots$ if $N$ is
not
      an integer or if $N$ is an integer and $n=0, 1, 2, \cdots , N$.
\item $\alpha_N^n = \beta_N^n$ if $n$ and $N$ are integers and $n>N$.
\end{itemize}
\par
For a positive integer $N$, it has been shown that the lowest $2N+1$
eigenvalues and the associated eigenvalues are exactly known as Lam\'e
polynomials, which are polynomials in terms of ${\rm sn}(u,k),
{\rm cn}(u,k)$ and ${\rm dn}(u,k)$. For general $N$, solutions of the
Lam\'e equation and even for integer $N$, other than $2N+1$
Lam\'e polynomials will not be written in such simple forms.
We explicitly present the Lam\'e polynomials for $N=1$ and $N=2$
below. The Lam\'e equation with $N=1$ has $2N+1=3$ Lam\'e
polynomials which are given,
apart from normalization constants, by
$$
\begin{array}{c|c}
\mbox{Lam\'e polynomial}& {\rm eigenvalue}\\ [0mm]\hline
Ec_1^0={\rm dn}(u,k) &   \alpha_1^0=k^2 \\ [2mm]
Ec_1^1={\rm cn}(u,k) &    \alpha_1^1=1 \\ [2mm]
Es_1^1={\rm sn}(u,k) &   \beta_1^1=1+k^2 \\[2mm]
\end{array}
$$
The Lam\'e equation with $N=2$ has $2N+1=5$ Lam\'e polynomials which are
give,
apart from normalization constants, by
$$
\renewcommand{\arraystretch}{1.3}
\begin{array}{c|c}
\mbox{Lam\'e polynomial}&{\rm eigenvalue}\\ [0mm]\hline
Ec_2^0={\rm sn}^2(u,k)-\frac{1+k^2+\sqrt{1-k^2+k^4}}{3k^2}
& \alpha_2^0=2(1+k^2-\sqrt{1-k^2+k^4}) \\ [2mm]
Ec_2^1={\rm cn}(u,k){\rm dn}(u,k)
& \alpha_2^1=1+k^2 \\ [2mm]
Es_2^1={\rm sn}(u,k){\rm dn}(u,k)
& \beta_2^1=1+4k^2 \\[2mm]
Es_2^2={\rm sn}(u,k){\rm cn}(u,k)
& \beta_2^2=4+k^2 \\ [2mm]
Ec_2^2={\rm sn}^2(u,k)-\frac{1+k^2-\sqrt{1-k^2+k^4}}{3k^2}
& \alpha_2^2=2(1+k^2+\sqrt{1-k^2+k^4}) \\[2mm]
\end{array}
$$
\par
Equipped with the eigenfunctions of the Lam\'e equation, we may expand
the fields $a_0, b_0, a_1$ and $b_1$ in (\ref{4reals})
in terms of these eigenfunctions. By taking account of the
boundary conditions
the bosonic fields $a_0$ and $b_0$ can be expanded
as (\ref{a0expan}) in the text.
The mass eigenvalue for each mode
$a_0^{(c,2n)}, a_0^{(s,2n+2)}, b_0^{(c,2n)}$ and $b_0^{(s,2n+2)}$ are
given by (\ref{a0emode}) in the text.
In the same way, by taking account of the boundary conditions,
the field $a_1$ and $b_1$ can be
expanded as (\ref{a1expan}) and (\ref{b1expan}), respectively.
The eigenvalues for the each
mode is given by (\ref{a1emode}) and (\ref{b1emode}).
\par
Let us consider the limit of
$k\rightarrow 0$ of the Lam\'e equation.
This limit corresponds to the limit of
$R\rightarrow R^*=1/{2\Lambda}$.
In this limit, the Lam\'e equation is reduced to
$-\frac{d^2}{du^2}\phi(u,k=0)=\Omega(k=0)\phi(u,k=0)$.
The boundary conditions in this limit become
$\phi(u+\pi)=\pm \phi(u)$.
The eigenvalues and the eigenfunctions are easily found, and they are
given by
$$
\begin{array}{c|cc}
{\rm eigenfunction}& {\rm eigenvalue}\\ \hline
Ec_N^0(u,k=0)=\frac{1}{\sqrt{2\pi R}}&
\alpha_N^0(k=0)=0 \\[2mm]
Ec_N^n(u,k=0)=\frac{1}{\sqrt{\pi R}}\cos(nu)&
\alpha_N^n(k=0)=n^2 &n=1, 2, 3\cdots \\[2mm]
Es_N^n(u,k=0)=\frac{1}{\sqrt{\pi R}}\sin(n u)&
\beta_N^n(k=0)=n^2& n=1,2,3 \cdots\ \\[2mm]
\end{array}
$$
Then, it follows that the expansions of the fields
$a_0(x,y), b_0(x,y), a_1(x,y)$ and $b_1(x,y)$ at $k=0$ become
\begin{eqnarray}
a_0(x,y)&=& \frac{1}{\sqrt{2\pi R}}a_0^{(0)}(x)+
\frac{1}{\sqrt{\pi R}}\sum_{n=1}^{\infty}
\Bigl[a_0^{(c,2n)}(x)\cos(\frac{ny}{R})
+a_0^{(s,2n)}(x)\sin(\frac{n y}{R})\Bigr],
\nonumber\\
b_0(x,y)&=& \frac{1}{\sqrt{2\pi R}}b_0^{(0)}(x)+
\frac{1}{\sqrt{\pi R}}\sum_{n=1}^{\infty}
\Bigl[b_0^{(c,2n)}(x)\cos(\frac{ny}{R})
+b_0^{(s,2n)}(x)\sin(\frac{n y}{R})\Bigr],
\nonumber\\
a_1(x,y)&=&\frac{1}{\sqrt{\pi R}}\sum_{n=1}^{\infty}
\Bigl[a_1^{(c,2n-1)}(x)\cos(\frac{(n-\half)y}{R})
+a_1^{(s,2n-1)}(x)\sin(\frac{(n-\half)y}{R})\Bigr],
\nonumber\\
b_1(x,y)&=&\frac{1}{\sqrt{\pi R}}\sum_{n=1}^{\infty}
\Bigl[b_1^{(c,2n-1)}(x)\cos(\frac{(n-\half)y}{R})
+b_1^{(s,2n-1)}(x)\sin(\frac{(n-\half)y}{R})\Bigr].
\label{a0expan1}
\end{eqnarray}
\vskip 0.5cm
\noindent{\bf B.3 Eigenvalue equation (\ref{fermieigen})}
\vskip 0.2cm

Let us consider the properties of the eigenvalues and associated
eigenfunctions of the equation
\begin{equation}
\Bigl[-\frac{d^2}{dy^2}+(U(y))^2+\frac{dU(y)}{dy}\Bigr]
\phi(y)=m_F^2\phi(y),
\label{48}
\end{equation}
where $U(y)\equiv 2k\omega{\rm sn}(\omega y,k)$ and
$\phi(y+4\pi R)=\phi(y)$.
Let $Ec^{\prime n}(u,k)$ and $Es^{\prime n}(u,k)$ be solutions of
the eigenvalue equation with
\begin{equation}
Ec^{\prime n}(u,k)\stackrel{k\rightarrow 0}{\longrightarrow} \cos(nu),
\quad
Es^{\prime n}(u,k)\stackrel{k\rightarrow 0}{\longrightarrow}\sin(nu),\\
\label{prop1}
\end{equation}
up to normalization constants.
Since the eigenvalues
of $Ec^{\prime n}(u,k)$ and $Es^{\prime n}(u,k)$
are found to be degenerate, as
we will prove later, it is not sufficient to specify $Ec^{\prime
n}(u,k)$
and $Es^{\prime n}(u,k)$ only by the conditions (\ref{prop1}).
The eigenfunctions $Ec^{\prime n}(u,k)$ and $Es^{\prime n}(u,k)$
may be supplemented by the condition of even/oddness under
the transformation $u\rightarrow -u$:
\begin{equation}
Ec^{\prime n}(-u,k) ~=~Ec^{\prime n}(u,k),\qquad
Es^{\prime n}(u,k) ~=~ -Es^{\prime n}(-u,k).
\label{prop2}
\end{equation}
Once we have two properties (\ref{prop1}) and (\ref{prop2}), the
eigenfunctions $Ec^{\prime n}(u,k)$ and $Es^{\prime n}(u,k)$
are uniquely specified up to normalization constants.
\par
Since the set of $\{Ec^{\prime n}(u,k),
Es^{\prime n+1}(u,k), n=0,1,2,\cdots\}$
is expected to form a complete set,
we can expand $\eta_{\pm}$ and $\zeta_{\pm}$ in terms of these
eigenfunctions as done in the text
(see (\ref{fexp1}) and (\ref{fexp2})).
Once we obtain the expansions for these spinor fields, the Dirac
spinors $\psi_i~(i=0,1)$ in $3$-dimensions are expanded as
\begin{eqnarray}
\!\!\!\!\!\!\!\!\!\!\!\!\psi_0(x,y)\!\!&\equiv& \!\!\frac{1}{\sqrt{2}}
(\chi_0(x,y)+i\rho_0(x,y))\nonumber\\
\!\!&=&\!\!\frac{1}{2}\sum_{n=0}^{\infty}\Bigl[
(\eta^{(c,n)}(x)+i\zeta^{(c,n)}(x))(Ec^{\prime n}(\omega y, k)+
Ec^{\prime n}(\omega(y+2\pi R),k))\nonumber\\
\!\!& &\!\!+(\eta^{(s,n+1)}(x)
+i\zeta^{(s,n+1)}(x))(Es^{\prime n+1}(\omega y, k)+
Es^{\prime n+1}(\omega(y+2\pi R),k))
\Bigr], \label{ngexp1}\\
\!\!\!\!\!\!\!\!\!\!\!\!\psi_1(x,y)\!\!&\equiv& \!\!\frac{i}{\sqrt{2}}
(\chi_1(x,y)+i\rho_1(x,y)) \nonumber \\
\!\!&=&\!\!\frac{i}{2}\sum_{n= 0}^{\infty}\Bigl[
(\zeta^{(c,n)}(x)+i\eta^{(c,n)}(x))(Ec^{\prime n}(\omega y, k)-
Ec^{\prime n}(\omega(y+2\pi R),k))\nonumber\\
\!\!& &\!\!+(\zeta^{(s,n+1)}(x)
+i\eta^{(s,n+1)}(x))(Es^{\prime n+1}(\omega y, k)-
Es^{\prime n+1}(\omega(y+2\pi R),k))
\Bigr].
\label{ngexp2}
\end{eqnarray}
\par
Let us discuss the relative magnitude of the eigenvalues $m_F^{(c,n)}$
and $m_F^{(s,n)}$ given by (\ref{hiefermi}) in the text.
The first and second relations in (\ref{hiefermi}) are easy to prove.
It is easy to see that at $k=0$ the eigenfunctions
$Ec^{\prime n}(u,k=0)$ and $Es^{\prime n}(u,k=0)$
have $2n$ nodes in the region $0\leq u < 4K(0)=2\pi$.
This fact implies that the number of modes of $Ec^{\prime n}(u,k)$
and $Es^{\prime n}(u,k)$ even for $0 \leq k < 1$ is equal to $2n$.
Since the number of nodes for $Ec^{\prime n}(u,k)$
or $Es^{\prime n}(u,k)$ is larger than that of
$Ec^{\prime m}(u,k)$ or $Es^{\prime m}(u,k)$ if $n>m$, we conclude
that $m_F^{(i,n)}> m_F^{(i, m)}~(i=c,s)$.
Let us next prove the third relation
$m_F^{(c, n)}=m_F^{(s,n)}$ if $n\neq 0$.
First, let us note that
\begin{equation}
\hat{a}^{\dagger}(u)\hat{a}(u)Ec^{\prime n}(u,k)=
(m_F^{(c,n)}/\omega)^2Ec^{\prime n}(u,k),
\label{opeqfermi}
\end{equation}
where
$\hat{a}(u)\equiv -\frac{d}{du}+2k~{\rm sn}(u,k)$ and
$\hat{a}^{\dagger}(u)\equiv \frac{d}{du}+2k~{\rm sn}(u,k)$.
Since ${\rm sn}(u+2K(k),k)=-{\rm sn}(u,k)$, $\hat{a}(u)$
and $\hat{a}^{\dagger}(u)$ satisfy
$\hat{a}(u+2K(k))=-\hat{a}^{\dagger}(u)$ and
$\hat{a}^{\dagger}(u+2K(k))=-\hat{a}(u)$.
It follows that
\begin{equation}
\hat{a}(u)\hat{a}^{\dagger}(u)Ec^{\prime n}(u+2K(k),k)=
(m_F^{(c,n)}/\omega)^2Ec^{\prime n}(u+2K(k),k).
\end{equation}
Multiplying the both sides of the above equation by
$\hat{a}^{\dagger}(u)$, we find that
$\hat{a}^{\dagger}(u)Ec^{\prime n}(u+2K(k),k)$ belongs to the
same eigenvalue $(m_F^{(c,n)}/\omega)^2$ as $Ec^{\prime n}(u,k)$.
We shall now show that the eigenfunction
$\hat{a}^{\dagger}(u)Ec^{\prime n}(u+2K(k),k)$
is indeed proportional to $Es^{\prime n}(u,k)$ if $n\neq 0$.
It is easy to show that
\begin{eqnarray}
\hat{a}^{\dagger}(u)Ec^{\prime n}(u+2K(k),k)
&\stackrel{u\rightarrow -u}{\longrightarrow}&
\hat{a}^{\dagger}(-u)Ec^{\prime n}(-u+2K(k),k) \nonumber\\
&=& -\hat{a}^{\dagger}(u)Ec^{\prime n}(u+2K(k),k).
\end{eqnarray}
Thus, the eigenfunction $\hat{a}^{\dagger}(u)Ec^{\prime n}(u+2K(k),k)$
is odd under $u\rightarrow -u$. We can further show that
\begin{eqnarray}
\hat{a}^{\dagger}(u)Ec^{\prime n}(u+2K(k), k)
&\stackrel{k\rightarrow 0}{\longrightarrow}&
\frac{d}{du}Ec^{\prime n}(u+\pi,k=0) \nonumber\\
&\propto& \sin(nu).
\end{eqnarray}
Therefore, the eigenfunction
$\hat{a}^{\dagger}(u)Ec^{\prime n}(u+2K(k),k)$
has been found to satisfy
the properties (\ref{prop1}) and (\ref{prop2}),
which $Es^{\prime n}(u,k)$ should satisfy,
and hence $\hat{a}^{\dagger}(u)Ec^{\prime n}
(u+2K(k),k)$ can be identified with
the eigenfunction $Es^{\prime n}(u,k)$,
up to normalization. It follows that $Ec^{\prime n}(u,k)$ and
$Es^{\prime n}(u,k)$ belong to the same eigenvalue, {\it i.e.}
$m_F^{(c,n)}=m_F^{(s,n)}$ if $n\neq 0$.

\newpage
\begin{center}
{\bf Table Captions}
\end{center}
\begin{center}
\begin{tabular}{c|cc} \hline
mode & {\it $({\rm mass})^2$} \\  \hline
$a_0^{(0)},b_0^{(0)}$  & 0 & \\
$a_0^{c(n)},a_0^{s(n)},b_0^{c(n)},b_0^{s(n)}$ &
$(\frac{n}{R})^2$, & $n\in{\bf Z}>0$\\
$a_1^{c(l)},a_1^{s(l)}$ & $|M|^2-\Lambda^2+
(\frac{l}{R})^2$,
& $l\in {\bf Z}+\half >0$ \\
$b_1^{c(l)},b_1^{s(l)}$ & $|M|^2+\Lambda^2+
(\frac{l}{R})^2$,
&  $l\in {\bf Z}+\half >0$ \\  \hline
\end{tabular}
\end{center}
\begin{center}
Table III-1. The boson mass spectrum
of the $Z_2$ model for $R \leq R^*$.
\end{center}
\par
\begin{center}
\begin{tabular}{c|cl} \hline
mode & {\it $({\rm mass})^2$} & \\  \hline
$\chi_0^{(0)},\rho_0^{(0)}$  & 0 & \\
$\chi_0^{c(n)},\chi_0^{s(n)},\rho_0^{c(n)},\rho_0^{s(n)}$ &
$(\frac{n}{R})^2$, & $n\in {\bf Z}>0$ \\
$\chi_1^{c(l)},\chi_1^{s(l)},\rho_1^{c(l)},\rho_1^{s(l)}$
& $|M|^2+(\frac{l}{R})^2$,
& $l\in {\bf Z}+\half >0$ \\  \hline
\end{tabular}
\end{center}
\begin{center}
Table III-2. The fermion mass spectrum
of the $Z_2$ model for $R \leq R^*$.
\end{center}
$$
\begin{array}{l|cl} \hline
\phi_{a_0}(u,k=0), \phi_{b_0}(u,k=0) & (m_{a_0})^2, (m_{b_0})^2\\
\hline
\phi_{a_0}^{(c,0)}, \phi_{b_0}^{(c,0)}=\frac{1}{\sqrt{2\pi R}}&0 \\[2mm]

\phi_{a_0}^{(c,2n)}, \phi_{b_0}^{(c,2n)}=\frac{1}{\sqrt{\pi
R}}\cos(2nu)&
4n^2\Lambda^2 &n\in{\bf Z}> 0 \\[2mm]
\phi_{a_0}^{(s,2n)}, \phi_{s_0}^{(s,2n)}=\frac{1}{\sqrt{\pi
R}}\sin(2nu)&
4n^2\Lambda^2 &n\in {\bf Z}>0\\ \hline
\phi_{a_1}(u,k=0) & (m_{a_1})^2 \\ \hline
\phi_{a_1}^{(c,2n-1)}=\frac{1}{\sqrt{\pi R}}\cos(2n-1)u&
((2n-1)^2-1)\Lambda^2 & n\in {\bf Z}>0 \\[2mm]
\phi_{a_1}^{(s,2n-1)}=\frac{1}{\sqrt{\pi R}}\sin(2n-1)u&
((2n-1)^2-1)\Lambda^2 & n\in {\bf Z}>0 \\ \hline
\phi_{b_1}(u,k=0)& (m_{b_1})^2\\ \hline
\phi_{b_1}^{(c,2n-1)}=\frac{1}{\sqrt{\pi R}}\cos(2n-1)u&
((2n-1)^2+1)\Lambda^2 & n\in {\bf Z}>0 \\[2mm]
\phi_{b_1}^{(s,2n-1)}=\frac{1}{\sqrt{\pi R}}\sin(2n-1)u&
((2n-1)^2+1)\Lambda^2 & n\in {\bf Z}>0 \\ \hline
\end{array}
$$
\begin{center}
Table III-3. The boson mass spectrum
of the $Z_2$ model in the $k\rightarrow 0$
limit.
\end{center}
\begin{center}
\begin{tabular}{c|c|c} \hline
${\rm eigenfunction}$ & $\Omega_{a_1}$ & $(m_{a_1})^2$  \\ \hline
$Ec_2^1(u,k)={\rm cn}(u,k){\rm dn}(u,k)$ & $\alpha_2^1(k)=1+k^2$ &$0$\\
$Es_2^1(u,k)={\rm sn}(u,k){\rm dn}(u,k)$ & $\beta_2^1(k)=1+4k^2$ &
$3k^2\omega^2$ \\
\hline
\end{tabular}
\end{center}
\begin{center}
Table III-4. Exact masses and eigenfunctions
of some lowest modes for $a_1$.
\end{center}
\begin{center}
\begin{tabular}{c|c|c} \hline
${\rm eigenfunction}$ &$\Omega_{b_1}$ &$(m_{b_1})^2$  \\ \hline
$Ec_1^1={\rm cn}(u,k)$ &$\alpha_1^1(k)=1$ &$(2+k^2)\omega^2$ \\
$Es_1^1={\rm sn}(u,k)$ & $\beta_1^1(k)=1+k^2$   & $2(1+k^2)\omega^2$
\\ \hline
\end{tabular}
\end{center}
\begin{center}
Table III-5. Exact masses and eigenfunctions
of some lowest modes for $b_1$.
\end{center}
\begin{center}
\begin{tabular}{c|c} \hline
$a_0, b_0$ & $({\rm mass})^2$  \\ \hline
${\rm lowest \ mode}$ &
$2k^2\omega^2$ + ${\cal O}(k^4)$ \\
${\rm 1st \ excited \ mode}$& $4\omega^2$+ ${\cal O}(k^4)$ \\
\hline
\end{tabular}
\end{center}
\begin{center}
Table III-6. Perturbative mass spectrum of the first two
lowest modes for $a_0$ and $b_0$.
\end{center}
\begin{center}
\begin{tabular}{c|c} \hline
$a_1, b_1$ & $({\rm mass})^2$ \\ \hline
${\rm 2nd\ excited\ mode\ of}\ a_1$ &
$(9-\frac{3}{2}k^2)\omega^2+{\cal O}(k^4)$  \\
${\rm 2nd\ excited\ mode\ of}\ b_1$ &
$(9-\frac{7}{2}k^2)\omega^2+{\cal O}(k^4)$ \\
\hline
\end{tabular}
\end{center}
\begin{center}
Table III-7.
Perturbative mass spectrum of the $2$nd excited modes
for $a_1$ and $b_1$.
\end{center}
\begin{center}
\begin{tabular}{c|cl} \hline
$\phi(u,k=0)$ & $(\frac{m_F(k=0)}{\Lambda})^2$ &  \\ \hline
$\cos(nu)$ & $n^2$ & $n\in{\bf Z} \geq 0$\\
$\sin(nu)$  & $n^2$ & $n\in {\bf Z}>0$\\
\hline
\end{tabular}
\end{center}
\begin{center}
Table III-8. Eigenvalues and eigenfunctions of (\ref{fermieigen})
in the limit of $k \rightarrow 0$.
\end{center}
\begin{center}
\begin{tabular}{c|cc} \hline
$\eta^{(c,n)},\eta^{(s,n)},\zeta^{(c,n)},\zeta^{(s,n)}$
& $({\rm mass})^2$  \\ \hline
$n=1~{\rm 1st \ excited \ state}$ &
$(1+\frac{13}{6}k^2)\omega^2$ + ${\cal O}(k^4)$  \\
$n=2~{\rm 2nd \ excited \ state}$ &
$(4+\frac{3}{40}k^2)\omega^2$ + ${\cal O}(k^4)$\\
$n=3~{\rm 3rd \ excited \ state}$ &
$(9-\frac{691}{280}k^2)\omega^2$ + ${\cal O}(k^4)$\\
\hline
\end{tabular}
\end{center}
\begin{center}
Table III-9. Perturbative mass spectrum of the higher modes of the
fermions
\end{center}
\begin{center}
\begin{tabular}{c|ccc} \hline
Symmetry&  $R \leq R^*$ & ${\it R > R^*}$&$R\rightarrow \infty$
\\ \hline
${\rm Supersymmetry}$& $\times$ & $\times$ &$\times$\\
${\rm Translational~inv.}$ & $\circ$ & $\times$ &$\times$\\
$Z_2$ & $\circ$ & $\times$ &$\times$\\  \hline
$U(1)_R$&$\times$ & $\circ$ & $\circ$ \\ \hline
\end{tabular}
\end{center}
\begin{flushleft}
Table III-10. Phase structure of the $Z_2$ model.
The $U(1)_R$ symmetry exists
only in the superpotential (\ref{equm}).
\end{flushleft}
\begin{center}
\begin{tabular}{c|ccc} \hline
Symmetry & $R \leq R^*$ & $R > R^*$ & $R\rightarrow \infty$\\
\hline
Supersymmetry  & $\times$ & $\times$ &$\circ$\\
Translational~inv. & $\circ$ & $\times$ &$\circ$\\
$U(1)$& $\circ$ & $\times$ &$\times$\\ \hline
$U(1)^{\prime}$ & $\circ$ & $\circ$ &$\circ$\\ \hline
$U(1)_R$ & $\times$ & $\circ$ & $\circ$ \\ \hline
\end{tabular}
\end{center}
\begin{flushleft}
Table IV-1. Phase structure of the $U(1)$ model.
The $U(1)_R$ symmetry exists
only in the superpotential (\ref{equp}) and the $U(1)^{\prime}$
symmetry is a linear combination of the translational invariance
and the $U(1)$ symmetry.
\end{flushleft}
\newpage
\begin{center}
\leavevmode{\epsfxsize=15cm\epsffile{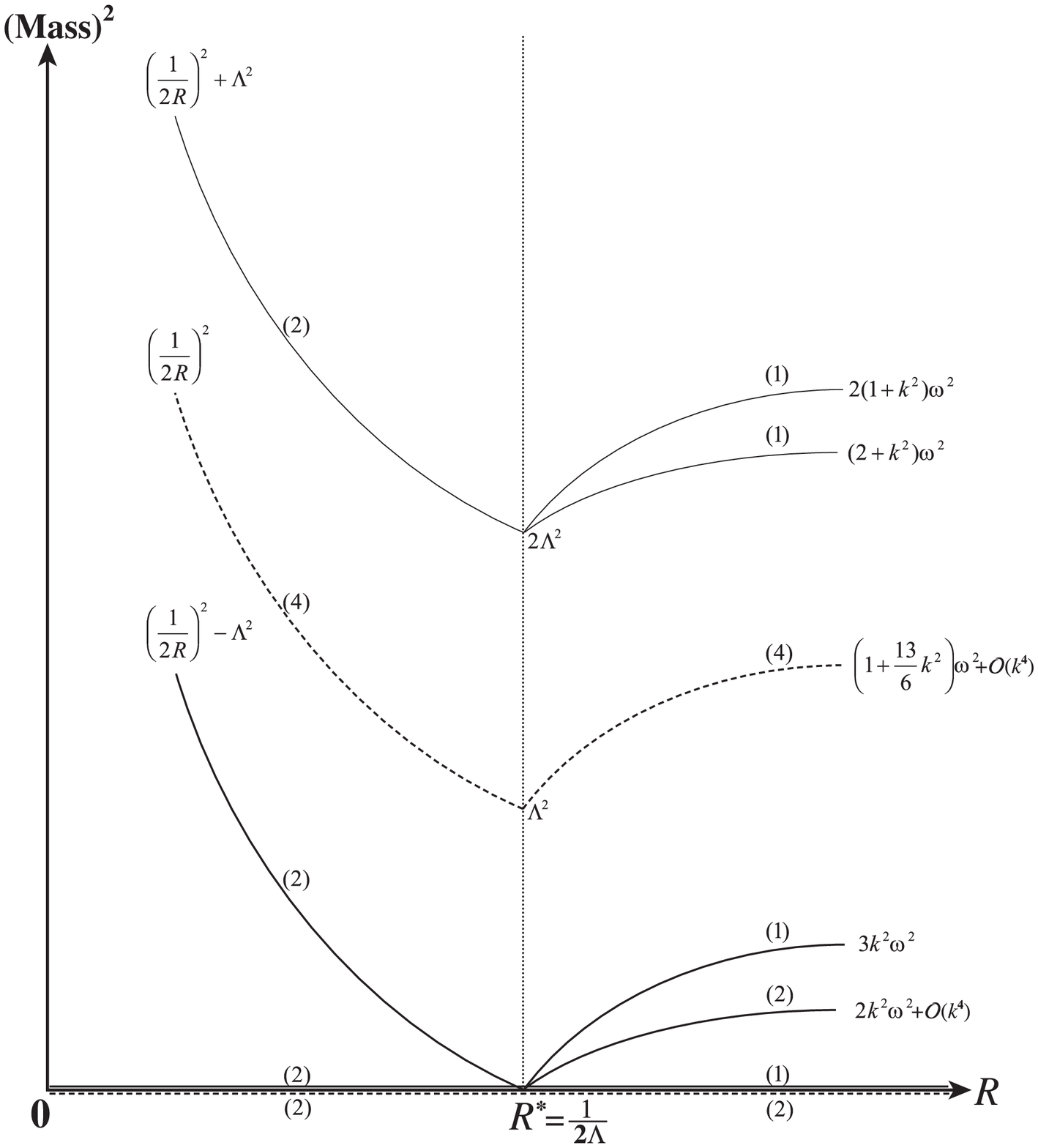}}
\end{center}
\smallskip
Figure 1: The $R$-dependence of the mass spectrum of the $Z_2$ model with
$<A_0>=\mu/g$ is depicted for a few lowest mass eigenstates.
The solid lines correspond to bosonic states and the dashed
lines to fermionic ones. The numbers in the parentheses represent
the degeneracy.
\newpage
\begin{center}
\leavevmode{\epsfxsize=15cm\epsffile{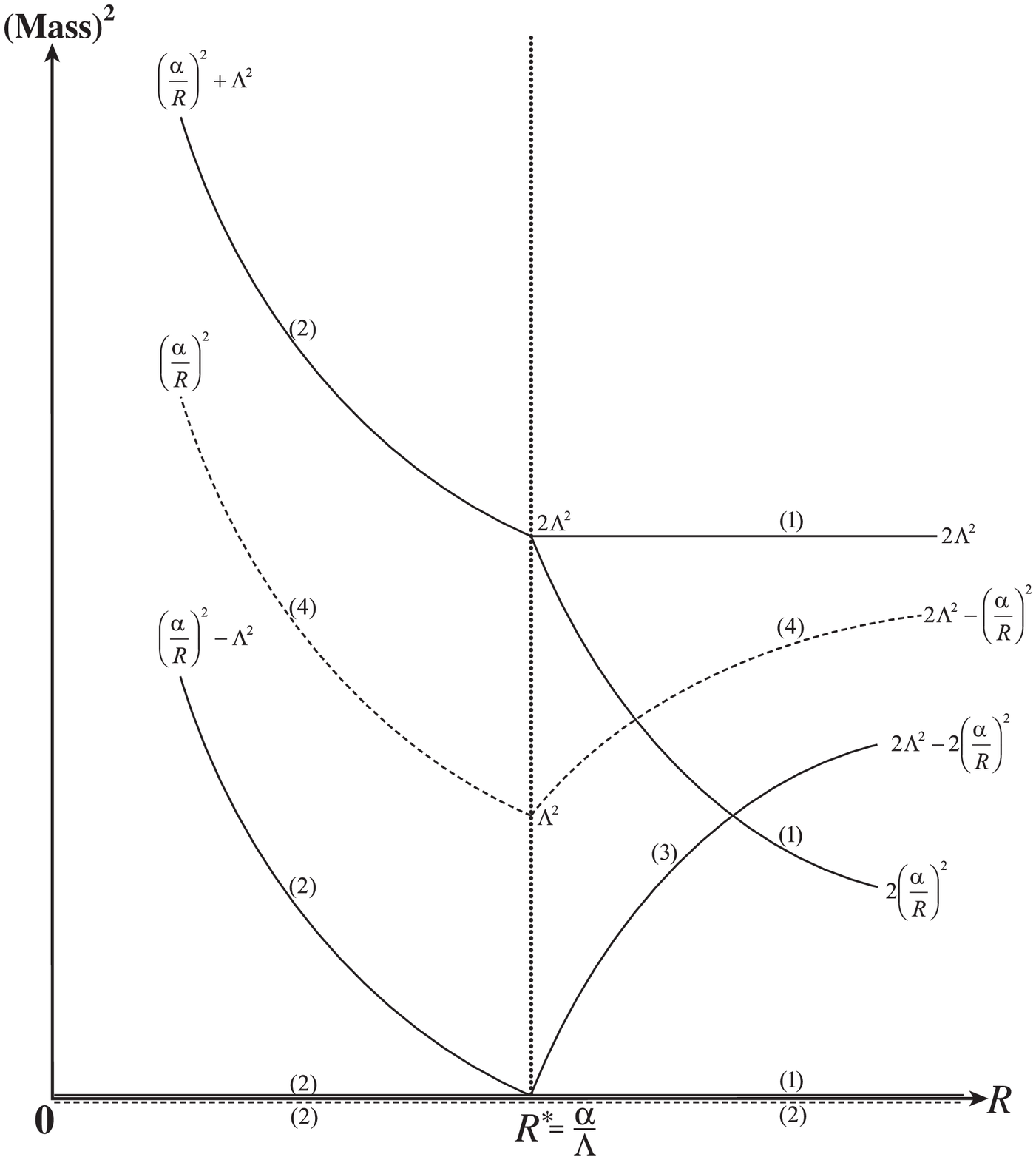}}
\end{center}
\smallskip
Figure 2: The $R$-dependence of the mass spectrum of the $U(1)$ model with
$\vev{A_0}=0$ is depicted for a few lowest mass eigenstates.
The solid lines correspond to bosonic states and the dashed
lines to fermionic ones. The numbers in the parentheses represent
the degeneracy.
\newpage
\begin{center}
\leavevmode{\epsfxsize=15cm\epsffile{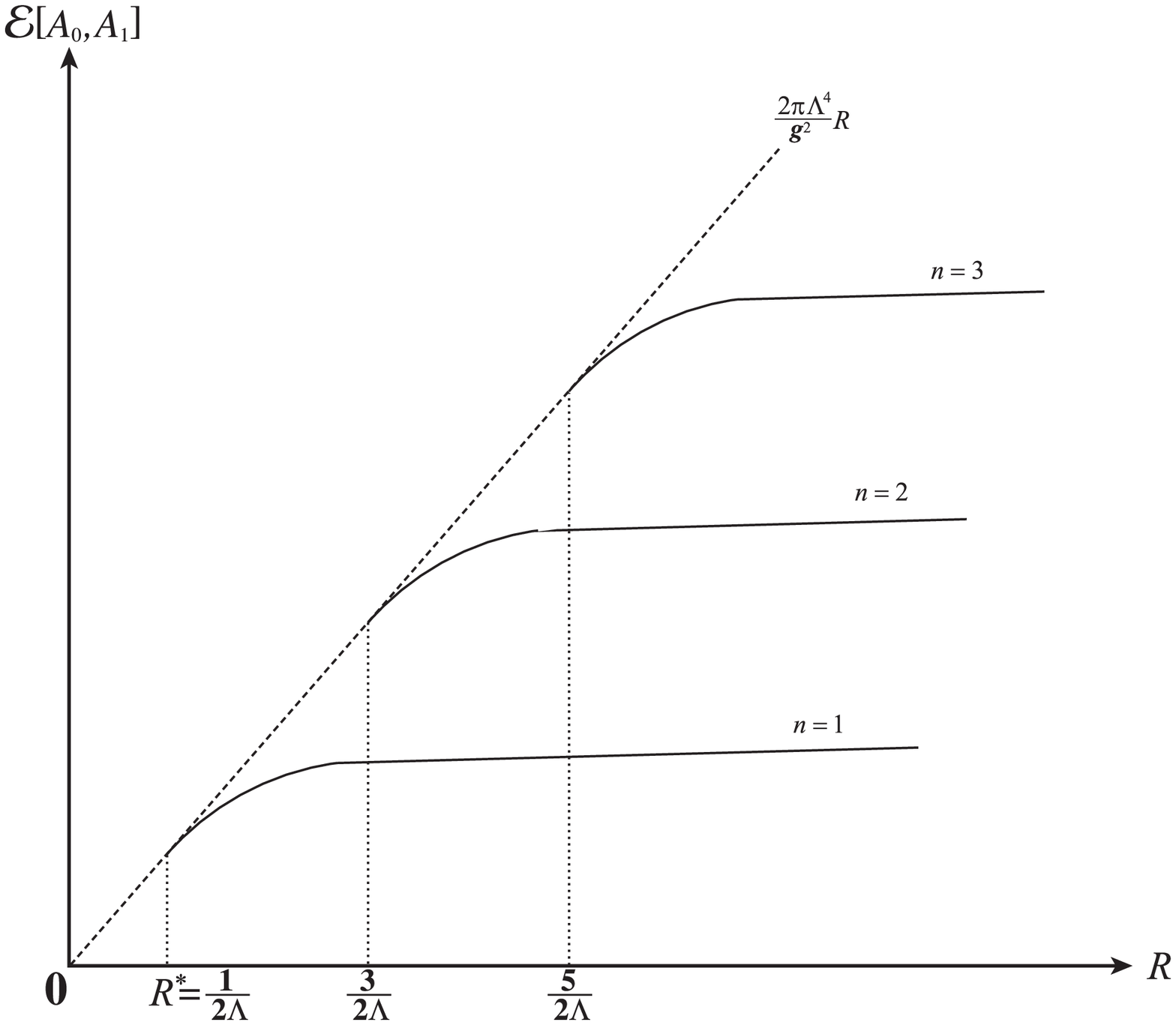}}
\end{center}
\smallskip
Figure 3: The $R$-dependence of ${\cal E}[A_0, A_1]$ in the $Z_2$ model.
The lines for $n=1, 2$ and $3$  correspond to 
${\cal E}[A_0^{({\rm II},1)},A_1^{({\rm II},1)}]$, 
${\cal E}[A_0^{({\rm II},2)},A_0^{({\rm II},2)}]$ and 
${\cal E}[A_0^{({\rm II},3)},A_0^{({\rm II},3)}]$, respectively.
We find the vacuum configuration for $R > R^*$ is given by
$A_0 = A_0^{({\rm II},1)}$ and $A_1 = A_1^{({\rm II},1)}$.


\begin{thebibliography}{99}
\bibitem{gsw} M. B. Green, J. H. Schwartz and E. Witten,~{\it
Superstring
Theory},~(Cambridge University Press, Cambridge, England, 1987).
\bibitem{kaluza} For a review,
see for example, D. Bailin and A. Love, \RPP{50}{87}{1087}.
\bibitem{ss} J. Scherk and J.H. Schwartz, \PLB{82}{79}{60}.
\bibitem{fi2} P. Fayet, \PLB{159}{85}{121}, \NPB{263}{86}{87}.
\bibitem{takenaga} K. Takenaga, \PLB{425}{98}{114},
\PRD{58}{98}{026004-1}.
\bibitem{hosotani} Y. Hosotani, \PLB{126}{83}{309}, \ANN{190}{89}{233}.
\bibitem{isham} C. J. Isham, \PRS{346}{78}{591}.
\bibitem{rs} K. D. Rothe and J. A. Swieca, \NPB{149}{79}{237}.
\bibitem{or} L. O'Raifeartaigh, \NPB{96}{75}{331}.
\bibitem{fi} P. Fayet and J. Iliopoulos.
\PLB{51}{74}{461}.
\bibitem{djackiw} Models which cause the spontaneous breakdown of
the translational invariance have been discussed in
Ref.\cite{jackiw1, odin}.
\bibitem{isham2} C. J. Isham, \PRS{363}{78}{581}. \\
V. B. Svetovoi and N. G. Khariton, \SJN{45}{87}{377}.
\bibitem{jackiw1} E. D'Hoker and R. Jackiw, \PRL{50}{83}{1719}.\\
E. D'Hoker, D. Z. Freedman and R. Jackiw, \PRD{28}{83}{2583}.
\bibitem{odin} S. D. Odintsov, Sov. J. Mod. Phys. {\bf 31} (1988) 695.
\bibitem{stt1} M. Sakamoto, M. Tachibana and K. Takenaga,
\PLB{457}{99}{231}.
\bibitem{stt2} M. Sakamoto, M. Tachibana and K. Takenaga,
\PLB{458}{99}{33}.
\bibitem{qdf} An interesting example of such mechanism has been found in
a special class of supersymmetric models \cite{qdeform}, in which
would-be supersymmetric
vacuum configurations have been removed from quantum moduli space due to
quantum deformed constraints.
\bibitem{qdeform} K. Izawa and T. Yanagida, \PTP{95}{96}{829}.\\
K. Intriligator and S. Thomas, \NPB{473}{96}{121}.
\bibitem{FGP} S. Ferrara, L. Girardello and F. Palumbo,
\PRD{20}{79}{403}.
\bibitem{shif} This observation is consistent with the Dvali-Shifman's
discussion \cite{ds} on how many SUSY is survived on the
field configuration with nontrivial coordinate dependence.
In our $Z_2$ model the central charge vanishes in
the kink background, so that the SUSY is completely broken
in $R\rightarrow \infty$ limit.
\bibitem{ds} G. Dvali and M. Shifman, \NPB{504}{97}{127}.
\bibitem{mu1} See also the discussions in the subsection {\bf 4.5}.
\bibitem{Lame} H. Bateman, Higher Transcendental Functions,
Vol.III (McGraw-Hill, New York, 1955).
\end{thebibliography}
\end{document}